\documentclass[rmp,twocolumn,superscriptaddress,showpacs,floatfix]{revtex4}

\usepackage{graphicx}
\usepackage{color}

\newcommand{\bce}{\begin{center}}

\newcommand{\beq}{\begin{equation}}
\newcommand{\eeq}{\end{equation}}
\newcommand{\bea}{\begin{eqnarray}}
\newcommand{\eea}{\end{eqnarray}}
\newcommand{\ave}[1]{\langle {#1} \rangle}

\newcommand{\ud}{\mathrm{d}}
\newcommand{\ket}[1]{| {#1} \rangle}
\newcommand{\bra}[1]{\langle {#1} |}

\begin{document}

\title{The Phase Diagram of Strongly Interacting Matter}

\author{P. Braun-Munzinger}
\affiliation{GSI Helmholtzzentrum f\"ur Schwerionenforschung mbH,
Planckstr, 1, D64291 Darmstadt, Germany}
\affiliation{Technical University Darmstadt, Schlossgartenstr. 9,
D64287 Darmstadt, Germany}
\author{J. Wambach}
\affiliation{GSI Helmholtzzentrum f\"ur Schwerionenforschung mbH,
Planckstr, 1, D64291 Darmstadt, Germany}
\affiliation{Technical University Darmstadt, Schlossgartenstr. 9,
D64287 Darmstadt, Germany}

\begin{abstract}
A fundamental question of physics is what ultimately happens to matter as it
is heated or compressed. In the realm of very high temperature and density
the fundamental degrees of freedom of the strong interaction, quarks and
gluons, come into play and a transition from matter consisting of confined
baryons and mesons to a state with 'liberated' quarks and gluons is
expected. The study of the possible phases of strongly interacting matter is
at the focus of many research activities worldwide. In this article we
discuss physical aspects of the phase diagram, its relation to the evolution
of the early universe as well as the inner core of neutron stars. We also
summarize recent progress in the experimental study of hadronic or
quark-gluon matter under extreme conditions with ultrarelativistic
nucleus-nucleus collisions.
\end{abstract}

\maketitle
\tableofcontents

\section{Introduction}

\begin{figure}[th]
\centerline{\includegraphics[width=0.45\textwidth]{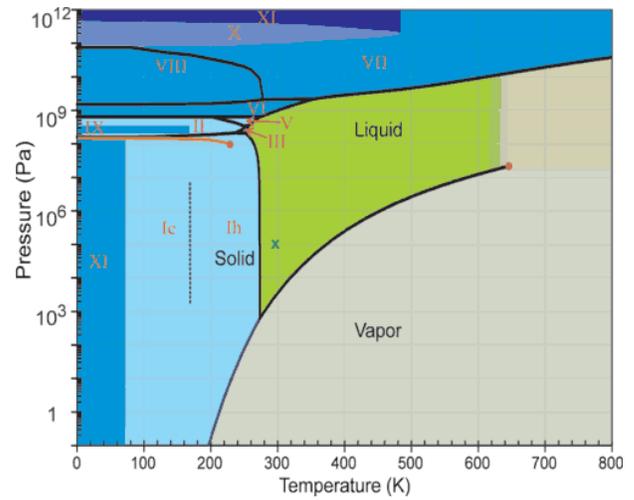}}
\vspace*{8pt}
\caption{The phase diagram of $H_20$~\cite{Chap}. Besides the liquid and 
gaseous phases a variety of crystalline and amorphous phases occur. Of
special importance in the context of strongly interacting matter is the
critical endpoint between the vapor and liquid phase.}
\label{phase_water}
\end{figure}

Matter that surrounds us comes in a variety of phases which can be
transformed into each other by a change of external conditions such as
temperature, pressure, composition etc. Transitions from one phase to another
are often accompanied by drastic changes in the physical properties of a
material, such as its elastic properties, light transmission, or electrical
conductivity. A good example is water whose phases are (partly) accessible to
everyday experience. Changes in external pressure and temperature result in a
rich phase diagram which, besides the familiar liquid and gaseous phases,
features a variety of solid (ice) phases in which the $\rm{H}_20$ molecules arrange
themselves in spatial lattices of certain symmetries (Fig.~\ref{phase_water}).

Twelve of such crystalline (ice) phases are known at present. In addition,
three amorphous (glass) phases have been identified. Famous points in the
phase diagram are the triple point where the solid, liquid, and gas phases
coexist and the critical endpoint at which there is no distinction between the
liquid and gas phase. This is the endpoint of a line of first-order liquid-gas
transitions; at this point the transition is of second order.


Under sufficient heating water and, for that matter any other substance, goes
over into a new state, a 'plasma', consisting of ions and free electrons. This
transition is mediated by molecular or atomic collisions. It is continuous
\footnote{Under certain conditions there may also be a true plasma phase
transition, for recent evidence see~\cite{fortov07}.}  and hence not a phase
transition in the strict thermodynamic sense. On the other hand, the plasma
exhibits new collective phenomena such as screening and 'plasma
oscillations'~\cite{thoma}.  Plasma states can also be induced by high
compression, where electrons are delocalized from their orbitals and form a
conducting 'degenerate' quantum plasma. In contrast to a hot plasma there
exists in this case a true phase transition, the 'metal-insulator'
transition~\cite{Mott,Gebhard}.  Good examples are white dwarfs, stars at the
end of their evolution which are stabilized by the degeneracy pressure of free
electrons~\cite{Chandra,Shapiro}.

One may ask what ultimately happens when matter is heated and compressed. This
is not a purely academic question but is of relevance for the early stages of
the universe as we go backwards in time in the cosmic evolution. Also, the
properties of dense matter are important for our understanding of the
composition and properties of the inner core of neutron stars, the densest
cosmic objects. Here, the main players are no longer forces of electromagnetic
origin but the strong interaction, which is responsible for the binding of
protons and neutrons into nuclei and of quarks and gluons into hadrons. In the
Standard Model of particle physics the strong interaction is described in the
framework of a relativistic quantum field theory called Quantum Chromodynamics
(QCD), where point-like quarks and gluons are the elementary constituents.


The question of the  fate of matter at very high temperature was first
addressed by Hagedorn in a seminal paper in 1965~\cite{Hage} and later
elaborated by Frautschi~\cite{Frautschi}. The analysis was based on the
(pre-QCD) 'bootstrap model' in which strongly interacting particles (hadrons)
were viewed as composite 'resonances' of lighter hadrons.  A natural
consequence of this model is the exponential growth in the density of mass
states
\beq
\rho(M_h)\propto M_h^{-5/2} e^{M_h/T_H}~.
\label{exp_hadron}
\eeq
This is well verified by summing up the hadronic states listed by the Particle
Data Group~\cite{PDG}. A fit to the data yields $T_H\sim 160-180$ MeV. It is
then easy to see that the logarithm of the partition function of such a 'resonance
gas' 
\beq
\ln {\cal Z}^{RG}(T)= \sum_{i}\ln {\cal Z}_i^{RG} +\kappa
\int_{m_0}^{\infty}\!\! dM_h~\rho(M_h) 
M_h^{3/2} e^{-M_h/T}
\eeq
and, hence, all thermodynamic quantities diverge when $T=T_H$, which implies
that matter cannot be heated beyond this 
limiting 'Hagedorn temperature'. Here, $\ln {\cal Z}_i$ is the logarithm of the
partition function for all well isolated particles with mass $m_i$. Above a
certain mass $m_0$ all particles start to overlap and from that point on the
sum is converted into an integral over the mass density $\rho(m)$ and all
particles can be treated in Boltzmann approximation. For the present argument
the explicit value of the constant $\kappa$ is immaterial.  The energy that is
supplied is used entirely 
for the production of new particles. This is of course at variance with our
present understanding of the big bang in which the temperature is set by the
Planck scale $T\sim M_{\rm Planck}=\sqrt{\hbar c/G_{N}}=1.22\times 10^{19}$
GeV where the 'Planck mass' is the mass for which the Schwarzschild radius is
equal to the Compton length divided by $\pi$. The quantity $G_N$ is the
Newtonian gravitational constant and $c$ is the speed of light\footnote{In all
formulas below we use $\hbar = c = 1$}. Referring to the Hagedorn paper
and the Friedman model of cosmology, Huang and Weinberg~\cite{HuangWein}
speculated in 1970 about a limiting temperature also in the big bang but noted:
{\it Our present theoretical apparatus is really inadequate to deal
with much earlier times, say when $T>100$ MeV.}

The situation changed in the early and mid 1970's after it  became clear
that hadrons are built from quarks and gluons and hence have substructure. 
In this context Itoh proposed in 1970 that there might exist stars that 
are entirely made of very massive quarks, rather than ordinary baryons~\cite{Itoh}\footnote{At
that time quarks were considered very heavy to account for the fact that no free
quarks were observed.}. The paradox of
Hagedorn was taken up in 1975~\cite{CaPa,CoPe} when it was noted that the
quark-gluon substructure of hadrons opened the possibility for a phase
transition to a new state of deconfined quark-gluon matter,  called
the 'quark-gluon plasma'\footnote{This term was coined by Edward
Shuryak~\cite{shuryak78}}. In close analogy to Fisher's droplet
model~\cite{Fisher_droplet} for phase transitions, Cabibbo and Parisi sketched
a very simple (second-order) phase boundary for the quark-hadron transition.
They argued that, when matter is sufficiently heated or compressed,
finite-size hadrons begin to overlap and quarks and gluons can travel freely
over large space-time distances. Within this picture, the limiting temperature
$T_H$ is in reality close to or even coincides with the critical temperature
for the phase transition between hadrons and
quarks and gluons. With point-like quarks and gluons the
temperature in the early universe can grow beyond bounds (big bang
singularity).

\section{Strongly Interacting Matter under Extreme Conditions}

\subsection{Quantum Chromodynamics}

To understand the salient features of the quark-hadron transition and to
appreciate the historical developments in its physical understanding we need
to recall some basic facts about the strong interaction. Its modern theory is
Quantum Chromodynamics, introduced in 1973~\cite{FGL}. This relativistic field
theory is formulated in close analogy to Quantum Electrodynamics (QED) as a
gauge theory of massive fermionic matter fields interacting with massless
bosonic gauge fields. In QED the Lagrangian density for the interaction of
electrons with photons is given by
\beq
{\cal L}_{QED}=-\frac{1}{4}F_{\mu\nu}F^{\mu\nu}+
\bar \psi\gamma^\mu i(\partial_\mu+ieA_\mu)\psi-m_e\bar \psi\psi
\label{LQED}
\eeq
where $F_{\mu\nu}$ denotes the field strength tensor of the electromagnetic
field, which in terms of the vector potential $A_\mu$ is obtained as 
\beq
F_{\mu\nu}=\partial_\mu A_\nu-\partial_\nu A_\mu~.
\eeq
The electrons are represented by the four-component Dirac spinor field $\psi$
of mass $m_e$ and the electric charge $e$ denotes the fundamental coupling
constant.  The Lagrangian is invariant under simultaneous (local) gauge 
transformations of the fermion field of the electron and the vector potential
\beq
\psi\to e^{-i\chi}\psi,\quad A_\mu\to A_\mu+\frac{i}{e}\partial_\mu\chi 
\label{gauge-u1}
\eeq
where $\chi(x)$ is a space-time dependent real valued function. The phase factor
$e^{-i\chi}$ is an element of the unitary group $U(1)$, which is hence called the 
'gauge group' of QED. 

Because of the smallness of
the fine structure constant $\alpha=e^2/4\pi \sim 1/137$ the evaluation of
physical processes can be carried out in perturbation theory with high
accuracy (the so calculated value for the magnetic moment of the electron agrees, e.g., to experiment
within ten 
decimals!). Historically this was one of the great triumphs of relativistic
field theories.

In QCD, quarks and gluons are the elementary degrees of freedom. Aside from the
relativistic quantum numbers dictated by Lorentz invariance, quarks come in
six 'flavors' (up, down, strange, charm, bottom, top). To obtain the correct
quantum statistics for hadronic wave functions it turns out that quarks as well
as gluons also have to carry 'color' as an additional quantum
number~\cite{Nambu,Greenberg}. The resulting Lagrangian density is then given by
\beq
{\cal L}_{QCD}=-\frac{1}{4}G^a_{\mu\nu}G_a^{\mu\nu}+\bar
q\gamma^\mu i(\partial_\mu+ig_s\frac{\lambda_a}{2}A^a_\mu)q-m_q\bar qq 
\label{LQCD}
\eeq
where $q$ includes the flavor and color quantum numbers to be appropriately
summed over. The 'strong coupling constant' $g_s$ is the analog of the electric
charge $e$ and $m_q$ denotes the quark mass of a given flavor. These masses
are generated in the electroweak sector of the Standard Model via the Higgs
mechanism, first introduced in the context of superconductivity. 
We will return to
this point and its physical implications later. The gauge
group structure is more complicated than in QED, since three colors are
required for each quark\footnote{Quarks form a fundamental representation of
the Lie group $SU(3)$.}. For group theoretical consistency also gluons, the
force carriers of the strong interaction, have to carry color charge (there are eight
vector potentials $A_\mu^a$ instead of one). As a physical consequence they will 
self-interact. Mathematically this is reflected by a modification
of the field strength 
tensor
\beq
G_a^{\mu\nu}=\partial^\mu A^\nu_a-\partial^\nu A^\mu_a-g_sf_{abc}A_b^\mu
A_c^\nu~,
\eeq
which now includes a non-linear term. Its form is entirely dictated by the
gauge group (which is now $SU(3)$ rather than $U(1)$) through its 'structure
constants' $f_{abc}$\footnote{Gauge groups other than $U(1)$ were first
discussed by Yang and Mills in 1954~\cite{YM} in the context of $SU(2)$ and
the corresponding field theories are therefore called 'Yang-Mills
theories'. Since the generators of $SU(N)$ do not commute, such theories are
also called 'non-Abelian'.}. The group 
structure is also reflected in the quark-gluon coupling through the
'Gell-Mann' matrices $\lambda_a$ which are the analog of the $SU(2)$ Pauli
matrices. Denoting the group 
of elements $SU(3)$ by $U(\chi^a)\equiv e^{-i\chi^a\frac{\lambda_a}{2}}$
and defining $A_\mu\equiv\frac{\lambda_a}{2}A_\mu^a$, the gauge transformation 
corresponding to Eq. (\ref{gauge-u1}) now reads
\bea
q\to U(\chi^a)q\nonumber\\
A_\mu\to U(\chi^a)A_\mu U^{-1}(\chi^a)+
\frac{i}{g_s}(\partial_\mu U(\chi^a))U^{-1}(\chi^a). 
\label{gauge-su3}
\eea
It obviously reproduces QED for the gauge group $U(1)$. 

The more elaborate group structure renders QCD much more complicated
than QED even at the classical level of Maxwell's equations.\footnote{For
instance, the wave equation for the vector potentials $A^a_\mu$ is
non-linear and its solutions in Euclidean space-time include solitons called
'instantons'.}

\begin{figure}[th]
\centerline{\includegraphics[width=0.40\textwidth]{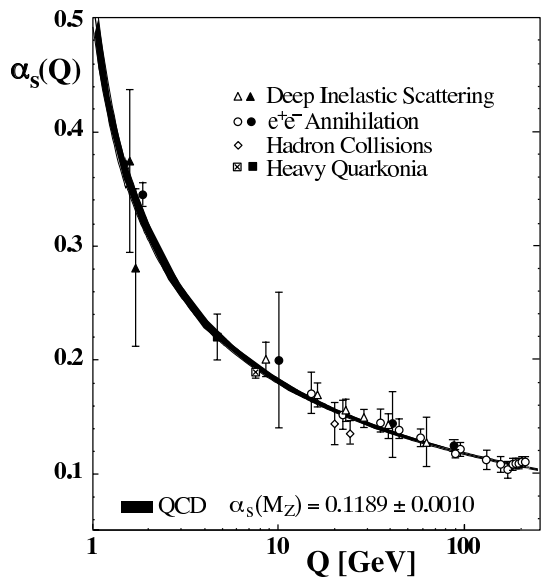}}
\vspace*{8pt}
\caption{The running of the fine structure constant of the strong interaction
with the momentum transfer $Q$ in a collision of quarks and/or
gluons~\cite{Bethke}.}  
\label{alpha_s} 
\end{figure}

In any relativistic field theory the vacuum itself behaves, due to quantum
fluctuations, like a polarizable medium. In QED the photon, although
uncharged, can create virtual electron-positron pairs, causing partial
screening of the charge of a test electron. This implies that the dielectric
constant of the QED vacuum obeys\footnote{Provided the distance
$r$ is large enough so that the virtual cloud around the test charge is not
penetrated. The distance scale is set by the inverse Compton 
wavelength of the electron, which is very small.} 
$\epsilon_0>1$. On the other hand, because
of Lorentz invariance, $\epsilon_0\mu_0=1$, i.e. the magnetic permeability
$\mu_0$ is smaller than unity. Thus the QED vacuum behaves like a diamagnetic
medium. In QCD, however, the gluons carry color charge as well as spin. In
addition to virtual quark-antiquark pairs, which screen a color charge and
thus would make the vacuum diamagnetic, the self-interaction of gluons can
cause a color magnetization of the vacuum and make it paramagnetic. This
effect actually overcomes the diamagnetic contribution from $\bar qq$ pairs
such that $\mu_0^c>1$. The situation is somewhat similar to the paramagnetism
of an electron gas, where the intrinsic spin alignment of electrons
overwhelms the diamagnetism of orbital motion. Since $\mu_0^c>1$ it follows
that $\epsilon_0^c<1$, so that the color-electric interaction between charged
objects becomes stronger as their separation grows ('infrared slavery'). In
this sense the QCD vacuum is an 'antiscreening' medium. As the distance $r\to
0$, on the other hand, $\mu_0^c$ and $\epsilon_0^c\to 1$, and the interaction
becomes weaker ('asymptotic freedom'). This gives rise to a pronounced
variation ('running') of the strong 'fine structure constant'
$\alpha_s=g_s^2/4\pi$ with (space-time) distance or momentum transfer $Q$. Its
mathematical form to leading order was worked out in 1973 by Gross and
Wilczek and independently by Politzer~\cite{GWP_1, GWP_2} and reads
\beq
\alpha_s(Q^2)=\frac{12\pi}{(33-2N_f)\ln
\left(Q^2/\Lambda^2_{\rm{QCD}}\right)};\quad Q^2\gg\Lambda^2_{\rm{QCD}}  
\label{alphas_run}
\eeq
where $\Lambda_{\rm{QCD}}\approx 200$ MeV is called the fundamental QCD scale
parameter. As indicated in Fig.~\ref{alpha_s}  
the running of $\alpha_s$ is by now confirmed by experiments to very high precision and the
authors were awarded the 2004 Nobel Prize in physics for their predictions.
Even though a mathematical proof is still missing, it is generally believed
that the strong increase of the coupling constant for low values of $Q$ is
responsible for the fact 
that isolated quarks and gluons have not been observed and are permanently
'confined' in composite hadrons.

\subsection{Models of the phase diagram}

A simple picture of confinement is provided by the MIT-Bag
model~\cite{MIT_bag}. Here, the idealized assumption is made that the QCD
vacuum is a perfect paramagnet with $\mu_0^c=\infty$ and $\epsilon_0^c=0$. A
hadron is formed by carving a spherical cavity (bag) with radius
$R\sim\Lambda_{\rm{QCD}}^{-1}\approx 1$ fm out of the physical vacuum. Inside the bag
the vacuum is trivial, i.e. $\mu_0^c=\epsilon_0^c=1$, and the interaction
between color charges is therefore weak. From the boundary conditions on the
chromoelectric and chromomagnetic fields it immediately follows that the color
fields are totally confined within the hadron.\footnote{The situation is
analogous to the case of a cavity in a perfect conductor (superconductor) 
with $\mu=0,\epsilon=\infty$ except that the role of $\mu$ and $\epsilon$
are interchanged.} The cost in energy density for creating the cavity is 
called the bag constant $B$. After filling the bag with three quarks for
baryons or quark-antiquark pairs for mesons and imposing appropriate boundary
conditions on the quark wave functions to prevent leakage of color currents
across the boundary, $B$ can be determined from a fit to known hadron masses.

For the quark-hadron transition the MIT-Bag model provides the following
picture: when matter is heated, nuclei eventually dissolve into protons and
neutrons (nucleons). At the same time light hadrons (preferentially pions) are
created thermally, which increasingly fill the space between the nucleons.
Because of their finite spatial extent the pions and other thermally produced
hadrons begin to overlap with each other and with the bags of the original nucleons
such that a network of zones with quarks, antiquarks and gluons is formed. At
a certain critical temperature $T_c$ these zones fill the entire volume in a
'percolation' transition. This new state of matter is the quark-gluon plasma
(QGP). The vacuum becomes trivial and the elementary constituents are weakly
interacting since $\mu_0^c=\epsilon_0^c=1$ everywhere. There is, however, a
fundamental difference to ordinary electromagnetic plasmas in which the
transition is caused by ionization and therefore gradual. Because of
confinement there can be no liberation of quarks and radiation of gluons below
the critical temperature. Thus a relatively sharp transition with $\Delta
T/T_c << 1$ is expected. We will return to this issue in the section on numerical solutions 
of QCD on a space-time lattice below. A
similar picture emerges when matter is strongly compressed. In this case the
nucleons overlap at a critical number density $n_c$ and form a cold degenerate
QGP consisting mostly of quarks. This state may be realized in the inner
core of neutron stars and its properties will be discussed later.

In the MIT-Bag model thermodynamic quantities such as energy density and
pressure can be calculated as a function of temperature and quark chemical
potential\footnote{In contrast to water, where the phase diagram is
usually characterized by pressure and temperature, the number density is
generally not conserved for relativistic systems. Therefore, the grand
canonical ensemble with state variables temperature and quark chemical
potential is used. For strong interactions $\mu_q$ ensures conservation of
baryon number and  $\mu_q > 0$ implies a non-vanishing net quark density 
$n_q$.}$\mu_q$
and the phase transition is inferred via the Gibbs construction of the phase
boundary. Under the simplifying assumption of a free gas of massless quarks,
antiquarks and gluons in the QGP at fixed $T$ and $\mu_q$ one obtains the
pressure 
\beq 
p_{QGP}(T,\mu_q)=37{\pi^2\over 90}T^4+\mu_q^2 T^2+{\mu_q^4\over 2\pi^2}-B~. 
\eeq         
To the factor $37=16+21$, 16 gluonic ($8\times 2$), 12 quark ($3\times2\times
2$) and 12 antiquark degrees of freedom contribute\footnote
{Here it has been assumed that
only up and down flavors contribute significantly to the quark pressure.}.  
For quarks an additional factor
of $7/8$ accounts for the differences in Bose-Einstein and
Fermi-Dirac statistics. The temperature dependence of the pressure follows a
Stefan-Boltzmann law, in analogy to the black-body radiation of massless photons.
The properties of the physical vacuum are taken into account by the bag constant $B$,
which is a measure for the energy density of the vacuum. By construction, the
quark-hadron transition in the MIT bag model is of first order, implying that the phase
boundary is obtained by the requirement that, at constant chemical potential,
the pressure of the QGP is equal to that in the hadronic phase. For the latter
the equation of state (EoS) of hadronic matter is needed. Taking for
simplicity a gas of massless pions of 3 different charge states, which yields 
$p_\pi(T,\mu_q)=(3\pi^2/90)T^4$, a simple
phase diagram emerges in which the hadronic phase is separated from the QGP by
a first-order transition line. Taking for the bag constant the original MIT
fit to hadronic masses, $B=57.5$ MeV/fm$^{3}$ one obtains $T_c\sim 100$ MeV at
$\mu_q=0$ and $\mu_c\sim 300$ MeV at vanishing temperature~\cite{Buballa}.

These results imply a number of problems. On the one hand, the transition
temperature is too small, as we have learned. We will come back to
this in the next section. On the other hand, at $3\mu_q=\mu_b\sim M_N$ (mass of
the nucleon $M_N=939$ MeV), where homogeneous nuclear matter consisting of
interacting protons and neutrons is formed, a cold QGP is energetically almost
degenerate with normal nuclear matter. Both problems are, however, merely of a
quantitative nature and can be circumvented by raising the value of $B$. More
serious is the fact that, at large $\mu_q$, a gas of nucleons because of its
color neutrality is always energetically preferred to the QGP. The biggest
problem is, however, that QCD has a number of other symmetries besides local
gauge symmetry which it shares with QED. Most notable in the present context
is chiral symmetry, which is exact in the limit of vanishing quark masses.
For massless fermions their spin is aligned either parallel (right handed) 
or antiparallel (left handed) to the momentum. Chirality of a massless fermion is a 
Lorentz-invariant concept, i.e. left(right)-handed particles remain left(right)-handed in 
all reference frames\footnote{At the same time massless left- and right-handed fermions 
transform into each other under a parity transformation.}. For
physical up and down quark masses of only a few MeV this limit is well
satisfied when comparing them to typical hadronic mass scales such as the mass of
the nucleon\footnote{Also the QED Lagrangian (\ref{LQED}) is chirally symmetric
in the limit of vanishing $m_e$. On atomic scales this symmetry is however
badly broken.}. Exact chiral symmetry implies that only quarks with the same
helicity or 'chirality' interact, i.e. the left-handed and
right-handed world completely decouple. This means in particular that physical
states of opposite parity must be degenerate in mass.

Similar to a
ferromagnet, where rotational symmetry is spontaneously broken at low
temperatures through spin alignment, also the chiral symmetry of the strong
interaction is spontaneously broken in the QCD vacuum as a result of the
strong increase of $\alpha_s$ at small momenta (Fig.~\ref{alpha_s}). Empirical
evidence is the absence of parity doublets in the mass spectrum of
hadrons. Since massless quarks flip their helicity at the bag boundary the
MIT-Bag model massively violates chiral symmetry. For 
the thermodynamic considerations discussed so far this is unimportant, but for
other aspects of the phase diagram chiral symmetry will be crucial.

There exist effective theories for the strong interaction which emphasize the
aspects of chiral symmetry and its spontaneous breaking in the physical
vacuum. One of the most thoroughly studied model in  connection with the
phase diagram dates back to early work by Nambu and Jona-Lasinio (NJL)
\cite{NJL_1, NJL_2} in 1961, i.e. before QCD was formulated. In its original
formulation the NJL model was a relativistic field theory for interacting
point-like nucleons of vanishing mass. When applied in the context of QCD, the
nucleons were later replaced by (nearly) massless up and down quarks and the
model Lagrangian takes the form
\beq
{\cal L}_{NJL}=
\bar q(i\gamma^\mu\partial_\mu-m_q)q+G\left[(\bar qq)^2+(\bar qi\gamma_5\vec
\tau q)^2\right]~. 
\label{LNJL}
\eeq 
The interaction between quarks and antiquarks is constructed in a
manifestly chirally invariant fashion such that ${\cal L}_{NJL}$ is invariant
under left-right transformations of the quark fields in the limit
$m_q\to 0$ (chiral limit)\footnote{Since two quark flavors are involved, the
transformation group is the direct product of the 'isospin group' $SU(2)$,
acting on left- and right handed quarks, i.e. $SU(2)_L\times SU(2)_R$.}. 
Gluons do not appear explicitly but are subsumed in
an effective short-range interaction of strength $G$ between the quarks. For
sufficiently large $G$, chiral symmetry is dynamically broken in the ground
state through the condensation of quark-antiquark pairs, i.e. the vacuum
expectation value $\ave{\bar qq}$ becomes non-vanishing. This is an effect
that cannot be produced by perturbation theory. As a consequence, a gap in the
quark energy spectrum occurs. This is in direct analogy to metallic
superconductivity in which, according to the Bardeen-Cooper-Schrieffer (BCS) \cite{bardeen}
theory, pairs of electrons interact via the exchange of lattice phonons and
condense. 

\begin{figure}[th]
\centerline{\includegraphics[width=0.30\textwidth]{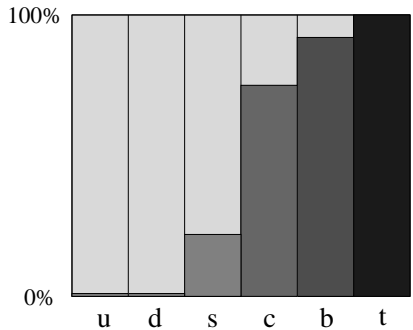}}
\vspace*{8pt}
\caption{Fraction of the effective quark mass generated dynamically
(light-grey) as compared to that from the Higgs mechanism in the electro-weak
sector of the Standard Model (dark grey).}  
\label{quark_masses}
\end{figure}

In a quantum field theory the elementary excitations of the vacuum
are interpreted as particles. In the original NJL model the energy gap
determines the mass of the nucleon. It is finite even in the absence of a
'bare mass'. Thus mass generation becomes intimately linked to the non-trivial
structure of the vacuum. In particle physics this idea of Nambu was new.
Replacing nucleons by quarks, the (nearly massless) quarks acquire a
'constituent mass' $M_q$ of around 300-400 MeV. Since a nucleon consists
essentially of three constituent quarks, its mass scale is thus explained. It
turns out that dynamical mass generation is not only a feature of the NJL
model but actually happens in QCD itself as can be shown from ab-initio
solutions of QCD at large coupling. Figure~\ref{quark_masses} summarizes the
current status of 
dynamical and Higgs contributions to the effective quark masses~\cite{Fischer}
using the Schwinger-Dyson formalism.

As can be seen, the dynamical contribution becomes less important the larger
the bare or Higgs mass of the quark. While the heaviest top-quark mass is
entirely generated by the Higgs mechanism, for up and down quarks close to 99
\% of their mass is dynamical. It is thus fair to say that almost all of the
mass in the visible universe is created through the non-perturbative structure
of the QCD vacuum.

In QCD, mesons emerge as bound states of quark-antiquark pairs with constituent
mass. Because of spontaneous chiral symmetry breaking there appears, however,
a peculiarity that is known from condensed matter physics and was first noted
by J. Goldstone \cite{Goldstone}. For vanishing (bare) quark mass there must
be a massless excitation of the vacuum, known as the 'Goldstone mode'. Such
highly collective modes occur e.g. in spin systems. The ferromagnetic ground
state has a spontaneous alignment of all spins. A spin wave of infinite
wavelength ($\lambda\to \infty, k\to 0$) corresponds to a simultaneous
rotation of all spins, which costs no energy\footnote{Spin waves obey the
dispersion relation $E\propto k^2$. In Lorentz-invariant theories $E\propto
k$ for massless particles.}. In strong interaction physics with two flavors,
this mode is identified with the pion. The fact that pions are not exactly
massless is related to the finite bare mass of the up and down quarks.
Nevertheless the pion mass with $\sim 140$ MeV is significantly smaller than
that of the $\rho$ or the $\omega$ meson ($\sim 800$ MeV $\sim 2M_q$).

\begin{figure}[th]
\centerline{\includegraphics[angle=-90,width=0.60\textwidth]{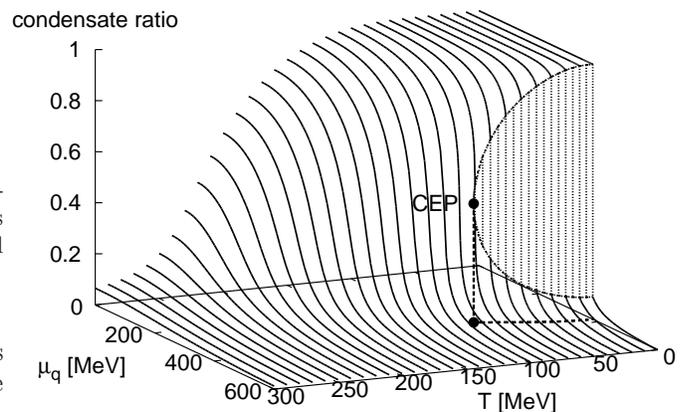}}
\vspace*{8pt}
\caption{Evolution of the chiral-condensate ratio $\ave{\bar
 qq}_{T,\mu_q}/\ave{\bar qq}$ plotted on the vertical axis with temperature
and quark chemical potential as predicted by the NJL
model~\cite{Heckmann}. The region of first order transitions, where the
condensate ratio jumps discontinuously, is clearly visible. The location of
the critical endpoint (CEP) and its projection onto the $T,\mu_q$-plane is
also indicated.}   
\label{qq_NJL}
\end{figure}

In the 1980's and 1990's the NJL model was used extensively in theoretical
studies of the phase diagram. Since it incorporates spontaneous symmetry
breaking and the ensuing mass generation, one can address questions of chiral
symmetry restoration with increasing $T$ and $\mu_q$ and the corresponding
medium modifications of hadron masses. The quark-antiquark condensate
$\ave{\bar qq}$ serves as an order parameter for chiral symmetry breaking,
analogous to the spontaneous magnetization in a spin system. Similar to the
Curie-Weiss transition, the order parameter vanishes at a critical
temperature $T_c$ in the chiral limit. This is the point where chiral symmetry
is restored and the quarks become massless\footnote{In the NJL model one
finds $M_q=m_q-2G\ave{\bar qq}$}. Figure~\ref{qq_NJL} displays a prediction for the evolution of the
chiral condensate with temperature and quark-chemical potential for physical
up and down quark masses obtained in mean-field theory.

While along the $T$-axis there is a continuous decrease indicating a smooth
restoration of chiral symmetry, one observes along the $\mu_q$-axis a first
order phase transition in which the condensate develops a discontinuity. With
increasing $T$ this transition becomes weaker and ends in a critical endpoint
(CEP) where the transition is second order. This is analogous to the
liquid-gas transition in water (Fig.~\ref{phase_water}).

\section{Results from Lattice QCD}

As described in the previous section one may predict, using the schematic bag
model and the NJL model 
(both focusing on different aspects of the strong interaction), that
upon heating and compression strongly interacting matter undergoes a
relatively abrupt transition from the hadronic phase to the QGP. The relevant
scales for this to happen are in the realm of very strong coupling
$\alpha_s\sim 1$. Hence, as for the description of any phase transition,  the
application of perturbative methods, which are 
very successful for QCD processes at high energies, must fail. The only known way to solve the
QCD equations from first principles in the region of strong coupling is to
discretize the QCD Lagrangian density  on a discrete
Euclidean space-time lattice. Here one makes use of the formal analogy between
Feynman's path-integral formulation of a quantum field theory in imaginary
time $\tau=it$ and the statistical mechanics of a system with temperature
$T=1/\tau$\footnote{The connection between a quantum system 
governed by the Hamiltonian $H$ and its statistical description
is made by considering the transition amplitude
$\bra{f}e^{-itH}\ket{i}$ from an initial state $i$ to the final state $f$. 
Comparing this to the partition function ${\cal Z}=\rm{Tr} (e^{-\beta H})$ ($\beta=1/T$)
one sees that $\cal{Z}$ can be obtained from the transition amplitude by the 
replacement $it=\beta$, setting $i=f=n$ and summing over $n$.}.
With this method of 'lattice QCD' the partition function of the
grand canonical ensemble  in the path integral formulation
\beq
{\cal Z}(V,T,\mu_q)=\int\!\!{\cal D}[A,q]\,e^{-\int_0^{1/T}d\tau\int_Vd^3x
\left({\cal L}^E_{\rm{QCD}}-i\mu_q q^\dagger q\right)}
\label{part_QCD}
\eeq
can be evaluated stochastically via Monte Carlo sampling of field
configurations, at least at vanishing $\mu_q$ (see below). 
In Eq.(\ref{part_QCD}) ${\cal L}^E_{\rm{QCD}}$ denotes the Euclidean version
of the QCD Lagrangian density (\ref{LQCD}). 

\begin{figure}[th]
\centerline{\includegraphics[width=0.45\textwidth]{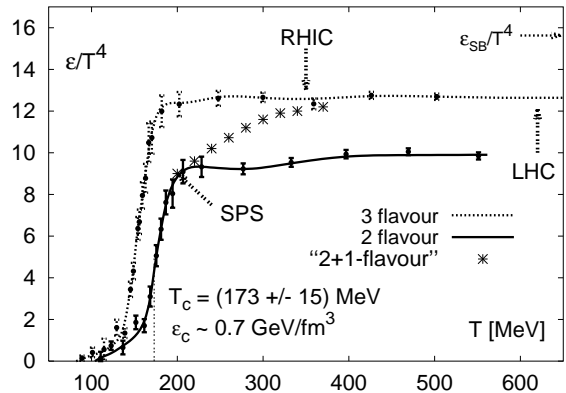}}
\vspace*{8pt}
\caption{The temperature dependence of the energy density from numerical
solutions of the QCD equations on a discrete (Euclidean) space-time lattice
\cite{karsch1}. The red and blue lines denote results with two quark flavors
(up and down) and three flavors (up, down, strange) of equal mass. The crosses
indicate the realistic case for which the strange quark mass is
roughly 150 MeV larger than the up and down masses. The arrows indicate the
corresponding energy densities and temperatures reached in current (SPS,
RHIC) and future (LHC) heavy-ion experiments (see sect. 4). For details see text.} 
\label{lattice_eos}
\end{figure}

From the partition function, the thermodynamic
state functions such as energy density and pressure can be determined as
\beq
\varepsilon\equiv \frac{E}{V}=\frac{T^2}{V}
\left(\frac{\partial\ln {\cal Z}}{\partial T}\right)_{V,\mu_q}
\!\!\!\!\!\!+\mu_q\frac{N}{V};\quad
p=T\left(\frac{\partial\ln {\cal Z}}{\partial V}\right)_{V,\mu_q}
\eeq
in the thermodynamic limit $V,N\to\infty; N/V=\rm{const}$.
At least for matter with an equal number of baryons and anti-baryons i.e. for
vanishing baryo-chemical potential $\mu_b=3\mu_q$ one obtains in this way
quantitative predictions for the temperature dependence of thermodynamic
quantities~\cite{karsch1}. Results are displayed in Fig.~\ref{lattice_eos}.

To emphasize deviations from the Stefan-Boltzmann behavior expected for a
free quark-gluon gas one typically shows the reduced energy density
$\varepsilon/T^4$ and pressure $p/T^4$. Near a critical temperature of
$T_c=175$ MeV the reduced energy density shows a rapid ($\Delta
T/T_c \leq 0.1$) variation, which
signals the transition from hadronic matter to the QGP. The critical energy
density $\epsilon(T_c)$ is $700\pm 300$ MeV/fm$^{3}$ which is roughly 5 times
higher than the energy density in the center of a heavy nucleus like
$^{208}\rm{Pb}$. At the same time the chiral condensate $\ave{\bar qq}=\partial
p/\partial m_q$ diminishes rapidly near $T_c$ signaling the restoration of
broken chiral symmetry. As indicated in Fig.~\ref{lattice_eos} a systematic
discrepancy of about 15 \% between the calculated energy density (and
pressure) and the free gas Stefan-Boltzmann limit is observed for $T>2T_c$.
Although this is roughly consistent with the first-order correction from perturbation
theory, the perturbation series is poorly convergent and resummation
techniques have to be employed~\cite{BIR} for a quantitative understanding of
the high-temperature EoS.

These ab-initio numerical findings support the simple model results for the
existence of a QGP transition discussed above. In this connection it should be
mentioned, however, that most lattice calculations still have to use
unrealistically large values for the light quark masses and rather small
space-time volumes. With anticipated high-performance computers in the range
of hundreds of Teraflop/s, these calculations will be improved in the near
future. Ultimately they will also provide definite answers concerning the
nature of the transition. Among others, this is of importance for primordial
nucleosynthesis, i.e. the formation of light elements, such as deuterium,
helium and lithium. In a strongly first-order quark-hadron transition, bubbles
form due to statistical fluctuations, leading to significant spatial
inhomogeneities. These would influence the local proton-to-neutron ratios,
providing inhomogeneous initial conditions for nucleosynthesis~\cite{Applegate,
Thomas93,DJSchwarz,Boyan}.
Other consequences would be the generation of magnetic fields, gravitational
waves and the enhanced probability of black-hole formation~\cite{Boyan}. 

At present, indications are that for $\mu_q=0$, relevant for the early
universe, the transition is a 'cross over', i.e. not a true phase transition
in the thermodynamic sense \cite{fodor_order}. Near $T_c$ the state functions
change smoothly but rapidly, as discussed above. For most of the
experimental observables to be discussed below this subtlety is, however, of
minor relevance.  A cross over would wash out large spatial fluctuations and
hence rule  out inhomogeneous cosmic scenarios. Very recent
studies~\cite{karsch_tc,fodor_tc} indicate that the exact value of the
transition temperature is still poorly known. In fact, these investigations
have yielded values for $T_c$ in the range 150 - 190 MeV. This is in part due
to difficulties with the necessary extrapolation to the thermodynamic
(infinite volume) limit and in part due to the general difficulty in providing
an absolute scale for the lattice calculations. Progress in this 
area is expected with simulations on much larger lattices at the next
generation computer facilities.

While at $\mu_q=0$ the lattice results are relatively precise, the ab-initio
evaluation of the phase boundary in the $(T,\mu_q)$-plane (Fig.~\ref{qq_NJL})
poses major numerical difficulties. This is basically related to the
Fermi-Dirac statistics of the quarks and is known in many-body physics as the
'fermion-sign problem'. For the integral (\ref{part_QCD}) this implies that
the integrand becomes an oscillatory function and, hence, Monte-Carlo sampling
methods cease to work. Only recently new methods have been
developed~\cite{Fodor2002,For2002,All2003,large_mu} to go into the region of
finite $\mu_q$.

What can be expected? Considering the phase boundary as a line of (nearly)
constant energy density, the bag model~\cite{bag_eos} predicts that the
critical temperature decreases with increasing $\mu_q$. By construction the
bag model describes a first-order phase transition for all chemical
potentials. For large values of $\mu_q$ and low temperatures there are
indications from various QCD-inspired model studies, chiefly the NJL model
(see Fig.~\ref{qq_NJL}), that the (chiral) phase transition is indeed first
order.  On the other hand, the lattice results discussed above seem to
indicate that, at very small $\mu_q$, the transition is a cross over. This would
imply that there is a critical endpoint in the phase diagram, where the line
of first-order transitions ends in a second-order transition (as in the
liquid-gas transition of water). In analogy to the static magnetic
susceptibility $\chi_M=\partial M/\partial H$ in a spin system one can define
a 'chiral susceptibility' as the derivative of the in-medium chiral condensate
$\ave{\bar qq}_{T,\mu_q}$ wrt the bare quark mass $m_q$ or equivalently as the
second derivative of the pressure, $\chi_m=\partial\ave{\bar
qq}_{T,\mu_q}/\partial m_q=\partial^2 p/\partial m_q^2$. Here the quark mass
$m_q$ plays the role of the external magnetic field $H$. In the Curie-Weiss
transition $\chi_M$ diverges. The same should happen with $\chi_m$ at the CEP.
On the other hand lattice studies and model calculations indicate that also
the quark number susceptibility $\chi_n=\partial n_q/\partial\mu_q=\partial^2
p/\partial \mu_q^2$ diverges. This implies that in the vicinity of the CEP the
matter becomes very easy to compress since the isothermal compressibility is
given by $\kappa_T=\chi_n/n_q^2$. It is conjectured that the critical behavior
of strongly interacting matter lies in the same universality class as the
liquid-gas transition of water~\cite{Stephanov}. The experimental
identification of a CEP and its location in the $(T,\mu_q)$ plane would be a
major milestone in the study of the phase diagram. Although very difficult,
there are several theoretical as well as experimental efforts underway
\cite{cpod06} to identify signals for such a point.  For a recent critical
discussion concerning the existence of a CEP in the QCD phase diagram
see~\cite{philipsen_07}.

\section{Experiments with Heavy Ions}

The phase diagram of strongly interacting matter can be accessed
experimentally in nucleus-nucleus collisions at ultrarelativistic energy, i.e.
energies per nucleon in the center of mass (c.m.) frame that significantly exceed
the rest mass of a nucleon in the colliding nuclei. After first intensive experimental
programs at the Brookhaven Alternating Gradient Synchrotron (AGS) and the CERN
Super Proton Synchrotron (SPS), the effort is at present concentrated at the
Relativistic Heavy-Ion Collider (RHIC) at Brookhaven. A new era of
experimental quark matter research will begin in 2009 with the start of the
experimental program at the CERN Large Hadron Collider (LHC). Here we will not
attempt to give an overview of the experimental status in this field (for
recent reviews see~\cite{pbm_js_nature,gyulassy_mclerran}) but concentrate on
a few areas which in our view have direct bearing on the phase diagram. Before
doing so we will, however, briefly sketch two of the key results from RHIC,
which have led to the discovery that quark-gluon matter in the vicinity of the
phase boundary behaves more like an ideal liquid rather than a weakly-interacting plasma.

\subsection{Opaque fireballs and the ideal liquid scenario}
At RHIC, Au-Au collisions are investigated at c.m. energies of 200 GeV per
nucleon pair. In such collisions a hot fireball is created, which subsequently
cools and expands until it thermally freezes out\footnote{A thermal freeze-out
is defined as the point in temperature where the density of particles with elastic 
cross section $\sigma$ becomes small
enough so that the mean free path $\lambda=1/n\sigma$ is larger 
than the system size.} and free-streaming hadrons reach
the detector.  The spectroscopy of these hadrons (and the much rarer photons,
electrons and muons) allow conclusions about the state of the matter inside
the fireball, such as  its temperature and density. The four experiments at
RHIC have recently summarized their
results~\cite{phenix,brahms,phobos,star}. For a complete overview see also the
proceedings of the three recent quark matter 
conferences~\cite{qm2005,qm2006,qm2008}. 

\begin{figure}[th]
\centerline{\includegraphics[width=0.45\textwidth]{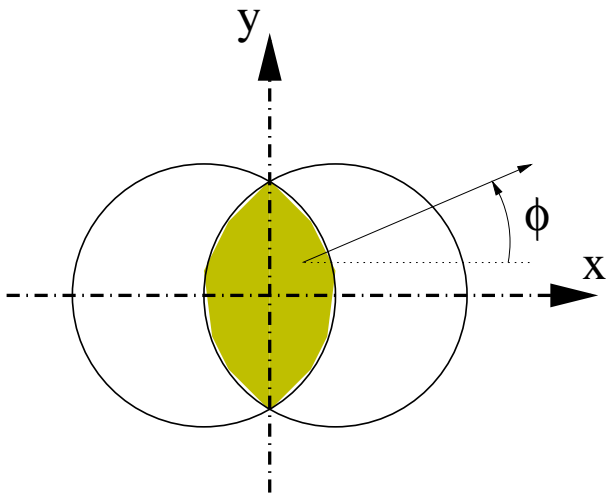}}
\vspace*{8pt}
\caption{Geometry of the
fireball in the plane perpendicular to the beam direction  in a
nucleus-nucleus collision with large  
impact parameter. }  
\label{almond}
\end{figure}

\begin{figure}[th]
\centerline{\includegraphics[width=0.45\textwidth]{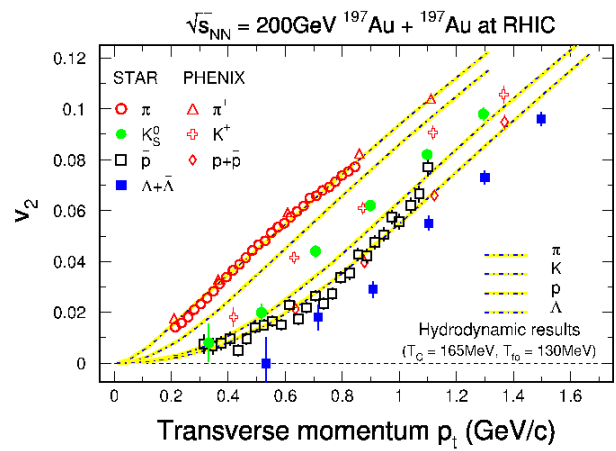}}
\vspace*{8pt}
\caption{The Fourier coefficient $v_2$ for pions, kaons, protons and $\Lambda$
baryons (with masses of 140, 495, 940 and 1115 MeV, respectively) emitted
with transverse momentum $p_t$ in semicentral Au-Au 
collisions at RHIC. The data are from 
the STAR collaboration~\cite{star_flow}. The lines correspond to
predictions~\cite{huovinen} 
from hydrodynamical calculations with an equation of state based on weakly
interacting quarks and gluons.}  
\label{hydro}
\end{figure}

The produced fireball has such a high density and temperature that apparently
all partons (quarks and gluons) reach equilibrium very rapidly (over a time
scale of less than 1 fm/c). Initially, the collision zone is highly anisotropic
with an almond-like shape, at least for collisions with not too small impact
parameter. The situation is schematically described in Fig.~\ref{almond}. In
this equilibrated, anisotropic volume large pressure gradients 
exist, which determine and drive the hydrodynamic evolution of the fireball. Indeed,
early observations at RHIC confirmed that the data on the flow pattern of the
matter follow closely the predictions~\cite{heinz_flow,shuryak_flow,huovinen}
based on 
the laws of ideal relativistic hydrodynamics. By Fourier analysis of the distribution
in azimuthal angle $\Phi$ (see Fig.~\ref{almond}) of the momenta of
produced particles, the Fourier coefficient $v_2 = \langle\cos(2\Phi)\rangle$
can be determined as a 
function of the particles transverse momentum $p_t$. These distributions can
be used to determine the anisotropy of the fireball's shape and are
compared, in Fig.~\ref{hydro} for various particle species, to the predictions
from hydrodynamical calculations. The observed close agreement between data and
predictions, in particular concerning the mass ordering of the flow
coefficients, implies that the fireball flows collectively  
like a liquid with negligible shear viscosity $\eta$.  Similar phenomena were
also observed in ultracold atomic gases of fermions in the limit of very large
scattering lengths, where it was possible, by measuring $\eta$ through
analysis of the damping rates of breathing modes, 
to establish that the system is in a strongly coupled
state~\cite{thomas}.

\begin{figure}[th]
\centerline{\includegraphics[width=0.45\textwidth]{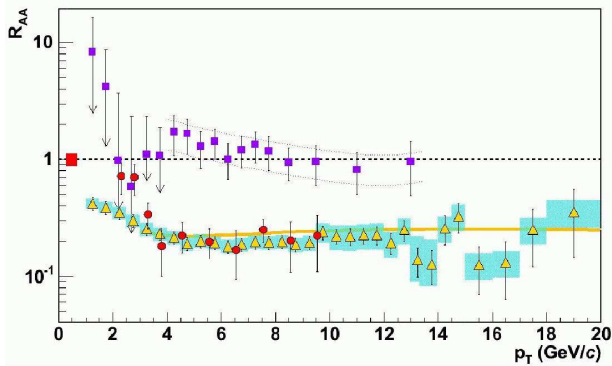}}
\vspace*{8pt}
\caption{Results from the PHENIX collaboration \cite{phenix_quench_1,phenix_quench_2} for the
$p_t$ dependence of the 
suppression factor $R_{AA}$. The suppression visible in the data for $\pi^0$
and $\eta$ mesons (yellow and red data points) provides evidence for the
presence of a dense medium 
scattering partons at high $p_t$ and degrading their momenta. Photons
(purple) which
undergo only electromagnetic interactions do not exhibit the effect. The
bands provide an estimate of systematic uncertainties. The yellow
solid line represents a theoretical spectrum \cite{vitev} calculated under
the assumption that the initially high $p_t$ parton loses energy by gluon
radiation in the dense gluon gas inside the fireball.}  
\label{quenching}
\end{figure}

This liquid-like fireball is dense enough that even quarks and gluons of high
momentum (jets) cannot leave without strong rescattering in the medium. This
'jet quenching' manifests itself in a strong suppression (by about a factor of
5) of hadrons with large momenta transverse to the beam axis compared to
expectations from a superposition of binary nucleon-nucleon collisions. The
interpretation is that a parton which eventually turns into a hadron must
suffer a large energy loss while traversing the hot and dense collision zone.
To make matters quantitative one defines the suppression factor $R_{AA}$ as
the ratio of the number of entries at a given transverse momentum $p_t$ in
Au-Au collisions to that in proton-proton collisions, scaled to the Au-Au
system by the number of binary nucleon-nucleon collisions such that, in the
absence of parton energy loss, 
$R_{AA} = 1$. Corresponding data are presented in
Fig.~\ref{quenching}. The strong suppression observed by PHENIX and, in
fact, by all RHIC collaborations \cite{phenix,phobos,brahms,star}
demonstrates the opaqueness of the fireball even for high momentum partons,
while photons, which do not participate in strong interactions, can leave the
fireball unscathed. Theoretical analysis of these data
\cite{vitev,gyulassy_mclerran} provides 
evidence, albeit indirectly, for energy densities exceeding 10 GeV/fm$^3$ in
the center of the fireball.
 Very interestingly, the fireball is apparently opaque enough
to strongly affect the spectra of heavy (c and b)
quarks~\cite{Adler:2005xv,Abelev:2006db}. This was not expected in view of
the arguments put forward in~\cite{Dokshitzer:2001zm}. Although the
mechanism for heavy quark energy loss is not well understood, the data
provide evidence for their scattering and thermalization in the
fireball. This will become important for the discussion about quarkonia below.
There is even evidence~\cite{Adler:2005ee} for the presence of Mach cone-like shock waves
\cite{machsto,machshu} caused by supersonic partons traversing the QGP.
Apparently both elastic parton-parton collisions as well as gluon radiation
contribute to the energy loss but it is fair to say that the details of this
mechanism are currently not well understood. The situation is concisely
summarized in~\cite{gyulassy_mclerran}. 

With the start of the nucleus-nucleus collision program at the
LHC in the Fall of 2009 the current understanding of jet-quenching and of
the ideal-fluid behavior of the hot fireball will be subjected to decisive
tests. At the much higher LHC energy, initial temperatures close to 1 GeV can
be reached and the fireball is probed with partons in the 100 GeV range. It
will be exciting to see how the currently developed concepts will evolve
with the data from this new era.

\subsection{Hadro-Chemistry}

\begin{figure}[th]
\centerline{\includegraphics[width=0.45\textwidth]{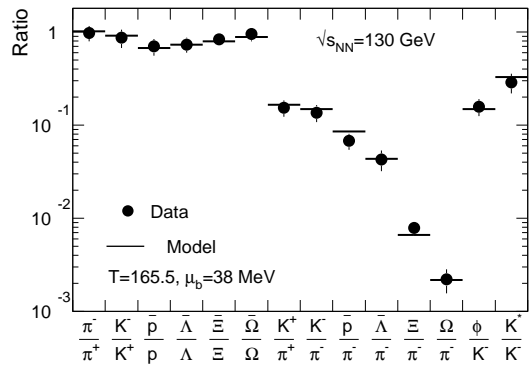}}
\vspace*{8pt}
\caption{Ratios of particle multiplicities in central Au-Au collisions at RHIC energies
in comparison with a fit~\cite{andronic} from a statistical model of
thermally and chemically equilibrated hadrons.}  
\label{ratios_ex}
\end{figure}

In ideal hydrodynamics no entropy is generated during the expansion and
cooling of the fireball, i.e. the system evolves through the phase diagram
along isentropes, starting in the QGP phase. This can be
experimentally verified through the production of a variety of mesons and
baryons. The analysis of particle production data at AGS, SPS and RHIC
energies has clearly demonstrated~\cite{andronic,bec1,bec2,review} that the
measurements can be understood to a high accuracy by a statistical ansatz in
which all hadrons are produced from a thermally and chemically equilibrated
state. This hadro-chemical equilibrium is achieved during or very shortly
after the phase transition and leads to abundances of the measured hadron
species that can be described by Bose-Einstein or Fermi-Dirac distributions
\beq
\label{hadron_mult}
n_j=\frac{g_j}{2\pi^2}\int^{\infty}_0
p^2dp\{\exp[(E_j(p)-\mu_j)/\rm{T}]\pm1\}^{-1} 
\eeq
of an ideal relativistic quantum gas. Here $E^2_j=M^2_j+\vec{p_j}^2$ is the
relativistic energy-momentum relation of free hadrons of mass $M_j$, $\mu_j$
the chemical potential of this species, and $g_j$ counts the number of degrees
of freedom, such as spin and charge state of a given hadron. The results of
such an analysis for the measured abundances in central Au-Au collisions at a
c.m. energy per nucleon pair of $\sqrt{s_{NN}}$ = 130 GeV at RHIC are shown in
Fig.~\ref{ratios_ex}.

\begin{figure}[th]
\centerline{\includegraphics[width=0.45\textwidth]{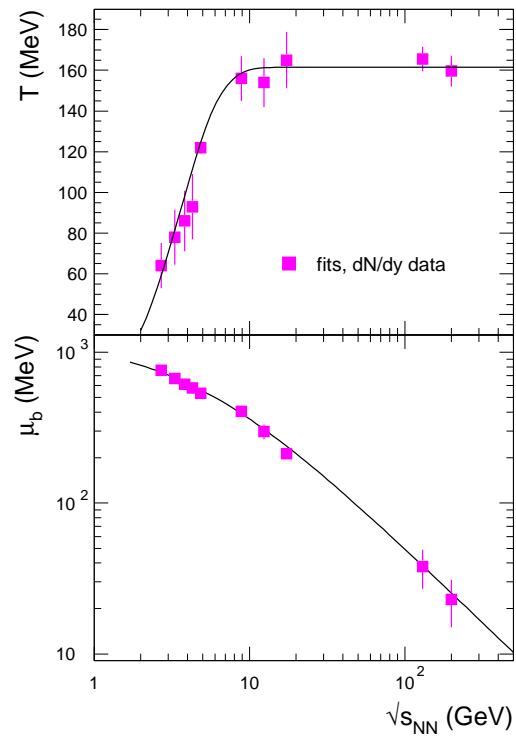}}
\vspace*{8pt}
\caption{Energy dependence of thermal parameters $T$ and $\mu_b$ 
from a statistical analysis~\cite{andronic} of hadrons produced in central
nucleus-nucleus collisions.}  
\label{tmus}
\end{figure}

Such calculations give, for each beam energy, a
set of two thermodynamic variables, namely temperature $T$ and baryo-chemical
potential $\mu_b$ at the point of hadroproduction, i.e. at chemical
freeze-out\footnote{Chemical freeze-out occurs, when inelastic collisions
 between 
particles cease such that the abundance ratios do not change anymore.}.
This is consistent with the assumption that all particles were
produced at the same instant, i.e. at 
the same temperature and chemical potential.  Such analyses also provide a
striking confirmation for the concept of a limiting temperature $T_H$
discussed above~\cite{Hage}, as shown in Fig.~\ref{tmus}.

\begin{figure}[th] 
\centerline{\includegraphics[width=0.45\textwidth]{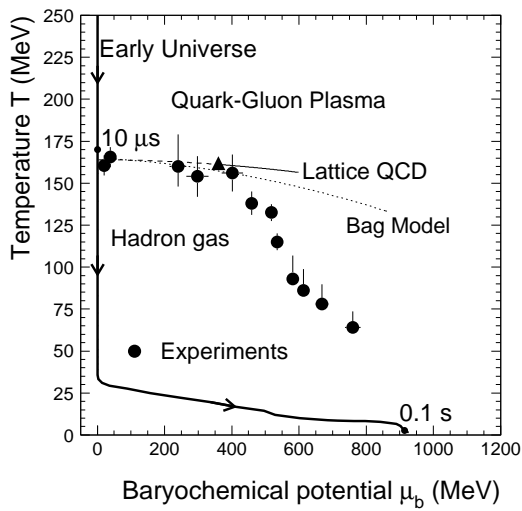}}
\vspace*{8pt}
\caption{The QCD phase diagram and data from a chemical freeze-out analysis in
nucleus-nucleus collisions at various beam energies. The data points closest
to the $T$-axis are from the highest collision energies. For the other
entries see text.}  
\label{phases_ex}
\end{figure}

The significance of these results is further appreciated by entering the
$(T,\mu_b)$ values of fixed beam energy into the phase diagram
(Fig.~\ref{phases_ex}), establishing the 'chemical freeze-out
curve'~\cite{pbm_sps,pbmjs_1,pbmjs_2}. It was noted early-on~\cite{cleymans_redlich}
that this curve can be understood phenomenologically by assuming that the
freeze-out takes place at a constant energy per particle of about 1 GeV.

In Fig.~\ref{phases_ex} the experimental points for chemical freeze-out are
compared with the phase boundary from lattice QCD \cite{fodor_cep} and from
the bag model \cite{bag_eos}. For illustration also a recent theoretical
prediction for the possible location of the CEP~\cite{fodor_cep} is shown
(triangle) as well as the trajectory that the early universe has taken in the
standard big bang model.\footnote{This trajectory is evaluated \cite{pbm_jw}
by assuming 
that the 
early universe expanded isentropically under the conditions of charge
neutrality and {net lepton - net baryon} number conservation and that the
entropy per baryon is fixed using the known baryon to photon ratio, see also
\cite{rafelski_02,kaempfer_univ}. Furthermore, the evolution proceeds in full 
chemical equilibrium among hadrons and leptons until the neutrinos freeze out
at a time of about t = 1 s.} It is interesting to note that, for $\mu_b<250$ MeV,
the experimental freeze-out points are close to the calculated phase boundary of
\cite{fodor_cep}. This does not come as a surprize.  On the contrary,
there are good arguments that the phase transition itself is responsible for
the equilibration of all hadron species. In a recent
analysis~\cite{pbm_js_wett} it was shown that, because of the strongly
increasing particle density near the phase boundary, multi-particle collisions
dominantly contribute to the particle production. This leads to a rapid
equilibration ($\tau < 1$ fm) of $\Lambda$ baryons and even of baryons with multiple strangeness
content ($\Xi,\Omega$ baryons) compared to the typical expansion time
of the fireball of several fm. It also explains naturally why all particles
freeze out in a relatively short time at nearly constant temperature.  Similar
conclusions, although based on different arguments, can be found in
\cite{stockpt,heinzpt,Heinz:2006ur}. The general result from these findings is that, at
least at small $\mu_b$, the temperatures extracted from the chemical analysis
are closely linked to $T_c$ obtained from the calculated QCD phase boundary.
Thus, for the first time, a fundamental parameter of the phase diagram, namely
the critical temperature $T_c$ at small $\mu_b$, has been confronted with
experiment and the agreement is very good\footnote{We neglect here the
above discussed uncertainty in $T_c$ obtained from recent lattice
calculations \cite{karsch_tc,fodor_tc}. It should be pointed out, however,
that a value of $T_c = 190 $~MeV is inconsistent with the scenario discussed
here and would probably 
imply the presence of an ultra-dense hadronic phase between chemical
freeze-out and the phase boundary. There is currently no indication of such
a phase from experiment.}.

For larger values of $\mu_b$ the measured freeze-out points deviate from the
predictions of lattice QCD (Fig.~\ref{phases_ex}). At present it is hotly
debated whether this deviation indicates the existence of a highly compressed
hadronic phase between the QCD phase boundary and the chemical freeze-out
line, or whether the calculation of the phase boundary at large $\mu_b$
will be modified by significant corrections from realistic quark masses and larger
space-time lattices. Important new insight is expected from 
measurements with the 'Compressed Baryonic Matter' (CBM) experiment planned at
the future 'Facility for Antiproton and Ion Research' (FAIR) at GSI in
Darmstadt, as well as from improved lattice simulations. 

The thermalization described above implies that equilibrated matter is
produced in high-energy collisions between nuclei.  In $e^+e^-$ and
hadron-hadron collisions, such an equilibration, in particular in the
strangeness sector, is not observed~\cite{review} although thermal
features are observed in the yields of produced particles~\cite{bec3,bec4}.
For very recent discussions of differences and similarities between $e^+e^-$
and nucleus-nucleus collisions see~\cite{andron_ee,bec5}.

For the high-energy domain accessible with Pb ions at the LHC
the scenario described implies essentially small changes in hadron
production (apart from an overall yield factor due to the much larger
volume). Any deviation would be a major surprize and would likely
indicate new physics. For speculations in this direction see~\cite{Rafelski:2008an}.

\subsection{Medium modifications of vector mesons}

As the spontaneously broken chiral symmetry of the strong interaction gets
restored at high temperatures and large chemical potentials, the quarks loose
their 'constituent' mass and only the 'bare' masses generated in the Higgs
sector of the Standard Model are left. As seen from Fig.~\ref{quark_masses}
this effect is most dramatic for up and down quarks and to a somewhat lesser
extent also for strange quarks. Most naively the mass of a hadron is a
multiple of the constituent quark mass $M_q$ (for baryons $M_b\sim 3M_q$ and
for mesons $M_m\sim 2M_q$) and one would therefore expect that all hadron
masses consisting of light u,d and s quarks should decrease significantly near
the phase boundary~\cite{Brown}. More general arguments along these lines led
to the conjecture of a general scaling law in which (nearly) all light hadrons
consisting of u,d quarks change with some power of the chiral condensate
ratio~\cite{BR} ('Brown-Rho scaling')\footnote{The pion is special because of
its 'Goldstone character' and therefore its mass should remain largely unaffected.}
\beq
M_h\propto (\ave{\bar qq}_{T,\mu_b}/\ave{\bar qq})^\alpha~.
\label{BR_scaling}
\eeq
Another obvious source of medium modifications of hadrons is the increased
collision rate in a hot and dense medium. As a consequence, many new decay
channels open, resulting in large widths. Finally, based on chiral symmetry
alone and its spontaneous breaking in the vacuum, it can be argued that the 
spectral properties of hadrons with opposite parity become more and more
similar as the chiral phase transition is approached.

Since possible modifications of hadron properties (masses, decay modes) occur
in the hot and dense phase of a heavy-ion collision, one needs an experimental
probe that is sensitive to this state of the matter. More than 30 years ago it
was suggested~\cite{Fein,Shur} that real or virtual\footnote{Virtual
time-like photons correspond to the process of di-lepton ($e^+e^-$ or 
$\mu^+\mu^-$) pair production or annihilation.} photons are ideal, since
they interact only electromagnetically with the surrounding matter and hence
leave the reaction zone almost undisturbed. Even at the highest temperatures
and compression reached in relativistic heavy-ion collisions the mean free
path of photons is typically $10^2-10^4$ fm, which is much larger than the
size of the fireball. 


Both longitudinal and transverse photon polarizations contribute to the
di-lepton rate, while real photons can only be transversely polarized. According to 
Fermi's Golden Rule the production cross section is directly related to the 
(auto)correlation function
$\ave{j_{elm}^\mu j_{elm}^\mu}$ of the electromagnetic current which involves the
charge carriers of the system. Taking quarks as fundamental constituents of
strongly interacting matter, $j_{elm}$ is given by
\bea
j_{elm}^\mu&=&\sum_{i=u,d,s}e_i\bar q_i\gamma^\mu q_i\nonumber\\
&=&\frac{2}{3}\bar u\gamma^\mu u-\frac{1}{3}\bar d\gamma^\mu d
-\frac{1}{3}\bar s\gamma^\mu s~. 
\eea
(For the measurements discussed below only the three light quark flavors are
relevant). It is well established by precision measurements that the $e^+e^-$
annihilation cross section below c.m. energies of $\simeq 1.2$ GeV is
essentially saturated by the light vector mesons $\rho,\omega,\phi$ with the
$\rho$-meson giving the largest contribution ($\sim 9:1:2$). Therefore, the
in-medium modification of the $\rho$-meson in di-lepton production in
heavy-ion collisions is of particular interest. Also the large ($\Gamma$ = 150
MeV) width implies that the $\rho$-meson decays and is regenerated several
times during the lifetime of the fireball: the resulting di-leptons then carry
information about its interior.

In physical terms the di-lepton signal is, therefore, dominantly due to pion
annihilation $\pi^+\pi^-\rightarrow \rho \rightarrow e^+e^-$ in the hadronic
phase or quark annihilation $\bar q q \rightarrow e^+e^-$ in the partonic
phase. If we assume that the fireball formed in an ultra-relativistic
nucleus-nucleus collision is close to thermal equilibrium then the above
formalism leads to di-lepton (photon) spectra after convolution of the
relevant transition rates  with the hydrodynamic
space-time evolution of the system.

Experiments to measure di-lepton production in nuclear collisions have been
conducted since the late 1980's starting with data taking at the DLS
experiment~\cite{dls1,dls2} in Berkeley. For a historical account of lepton
pair production measurements in general see~\cite{specht_tokyo}. Here we focus
on the most recent measurements at ultra-relativistic energies and the current
status of their interpretation.  To search for non-trivial contributions the
data for di-lepton measurements are compared to predictions for yields
resulting from the electromagnetic decay of hadrons at chemical freeze-out.
The hadron production rate is either directly measured or inferred from
statistical model calculations discussed above~\cite{review}. The resulting
yields are called 'hadronic cocktail' as they result from the standard
known mixture of unmodified hadronic resonances.

Pioneering results on the production of $e^+e^-$-pairs came from the DLS~\cite{dls2},
HELIOS~\cite{helios}, and CERES ~\cite{ceres_95,ceres_98,ceres_9596}
collaborations: the main and dramatic outcome of these experiments was that all
central nucleus-nucleus collision measurements exhibited a yield
that is strongly enhanced compared to predictions for cocktail decays in the
invariant mass range $0.2 < m_{e^+e^-} < 1.1$ GeV.  Theoretical analysis of
the excess observed in the CERES data~\cite{RaWa} indicated that the
enhancement is due to a strong increase of the $\rho$-meson width in the hot
and dense medium formed in the collision. The excess disappears for more
peripheral collisions~\cite{ceres_9596} (which exhibit features more like
nucleon-nucleon collisions) but for SPS energies the beam energy dependence of the observed
effect is small~\cite{ceres_03}. A satisfactory explanation of the excess
observed by DLS at much lower energies remained missing.

\begin{figure}[th] 
\centerline{\includegraphics[width=0.45\textwidth]{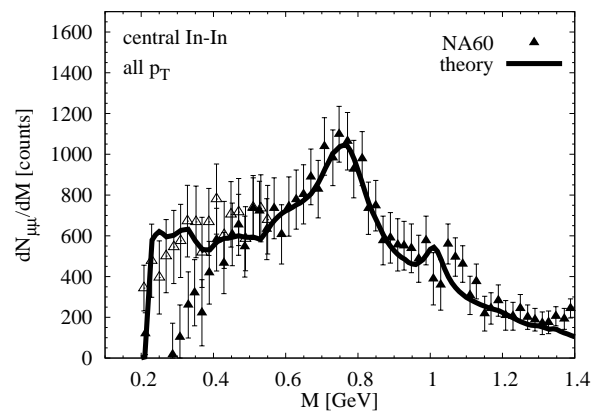}}
\vspace*{8pt}
\caption{The NA60 low mass spectrum after subtraction of 'cocktail' decay
processes after thermal freeze-out compared with the theoretical predictions
from refs.~\cite{rapp_hees1,rapp_hees2}. The open data points exhibit the
size of the correction resulting from a decrease in the $\eta$ yield by 10\%
\cite{na60_1}.}  
\label{na60_lowmass}
\end{figure}

Dramatic progress was recently achieved by the NA60 collaboration which,
for a collision system of intermediate mass (In+In), provided
data~\cite{na60_1} in the 
di-muon channel with very good statistics and improved mass resolution
compared to previous measurements. The quality of the data is
such that the di-lepton yield resulting from final state hadron decays,
i.e. the cocktail yield, can be subtracted from the measured di-lepton
spectra. The resulting subtracted spectrum is compared, in
Fig.~\ref{na60_lowmass}, with predictions that take into account all collision 
processes of the $\rho$ meson with the surrounding particles in the
fireball~\cite{rapp_hees1,rapp_hees2}.  
Note that the data are not acceptance corrected, implying that the
calculations have to be filtered appropriately for a meaningful
comparison~\cite{specht_tokyo,na60_1}.

In such many-body calculations, the spectral function of the $\rho$ meson is
considerably broadened in the hot and dense medium compared to the line shape
of the $\rho$ meson in vacuum. The strong broadening is dominantly due to
interactions of $\rho$ mesons with baryons (and anti-baryons) in the dense
fireball near the phase boundary. This is indeed observed in the NA60 data, as
is demonstrated by the quantitative agreement between data and calculations.
Note that there is no evidence for a possible downward shift of the
$\rho$ mass, as had been predicted early-on~\cite{BR} based on
a scaling relation (\ref{BR_scaling}) between the $\rho$ mass and the in-medium
quark condensate.

\begin{figure}[th] 
\centerline{\includegraphics[width=0.40\textwidth]{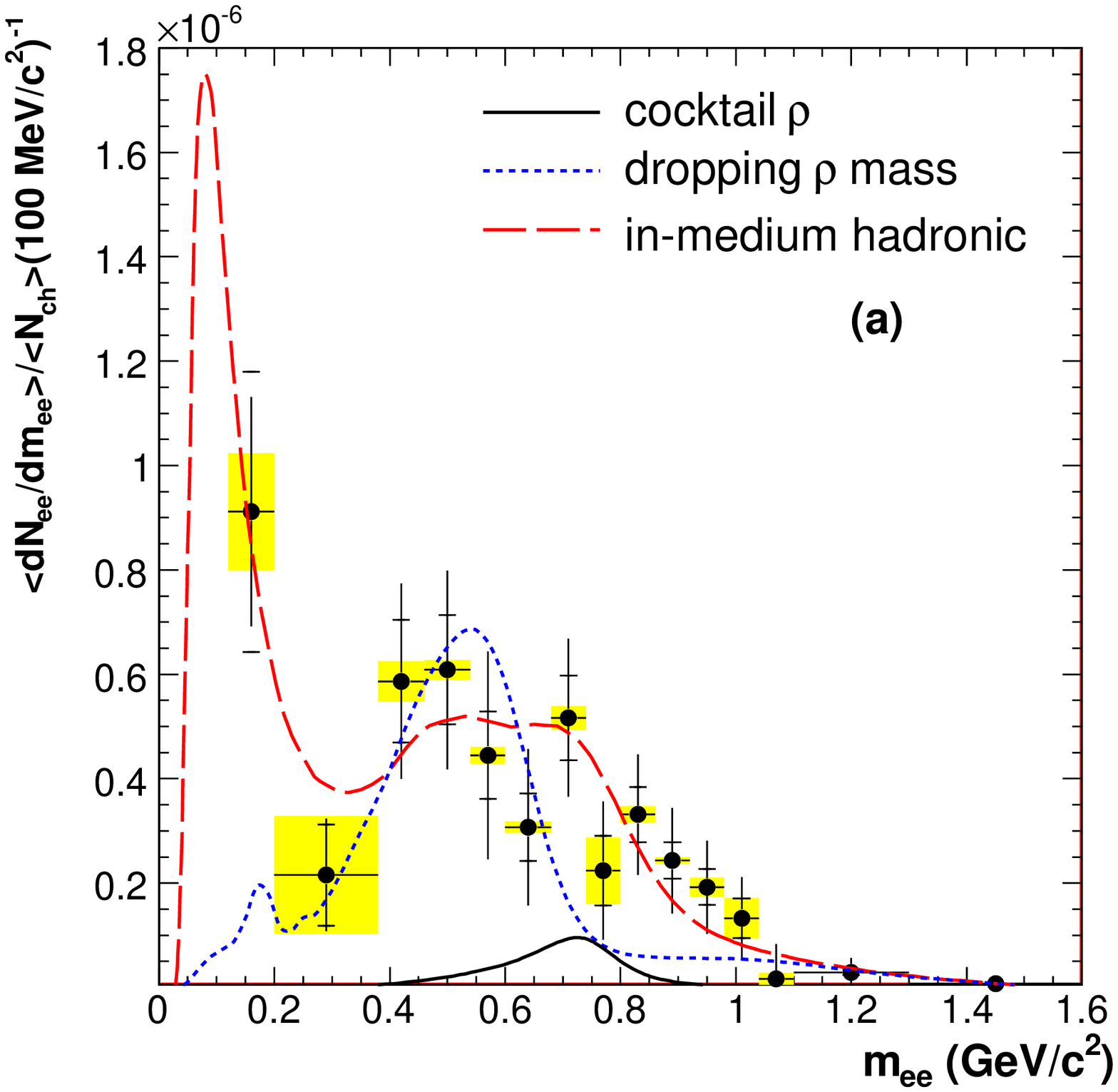}}
\vspace*{8pt}
\caption{The CERES low mass spectrum compared with theoretical predictions
from the 'dropping mass' scenario~\cite{BR} and hadronic many-body
theory~\cite{RaWa,rapp_hees2,Rapp_private}.}  
\label{ceres_lowmass}
\end{figure}

The CERES collaboration has recently presented~\cite{ceres_06} their
absolutely normalized data on
low-mass $e^+e^-$-pair production in central Pb-Au collisions at SPS energy,
taken with the upgraded CERES apparatus. Again, to explicitly display the
shape of the in-medium contribution to the di-lepton mass spectrum, the
cocktail excluding the $\rho$-meson contribution was subtracted from both the
data and theoretical calculations. The result is shown in~Fig.~\ref{ceres_lowmass}.

Note that the yield of the 'cocktail $\rho$' is only about 10 \% of the
observed yield near the mass of the vacuum-$\rho$ , demonstrating the extremely strong
modification of its spectral function in the dense fireball. Calculations
based on the many-body approach of~\cite{RaWa,rapp_hees2,Rapp_private}
explain this medium modification quantitatively, while those based on a
downward shift of the $\rho$ mass~\cite{BR,Rapp_private} are at variance with
the observations\footnote{See~\cite{BR3} for an updated view on the connection
between the $\rho$ mass and the chiral condensate.}, in accord with the
findings of the NA60 collaboration. Below di-lepton masses of 200 MeV, the CERES
data indicate  a further strong rise. Such an increase
towards the 'photon point' ($m_{e^+e^-} = 0$) was predicted in a consistent
treatment of the in-medium $\rho$-meson spectral function~\cite{RaWa} and its
observation lends further support to the underlying theoretical approach.

Two more experiments have released data on di-lepton production in
nucleus-nucleus collisions during the past year. The HADES collaboration
presented their first data on C+C collisions at relatively low
energy~\cite{hades1,hades2}, substantially corroborating the measurements of
the DLS collaboration. Currently a significant theoretical effort is underway
to understand these observations. 

\begin{figure}[th] 
\centerline{\includegraphics[width=0.40\textwidth]{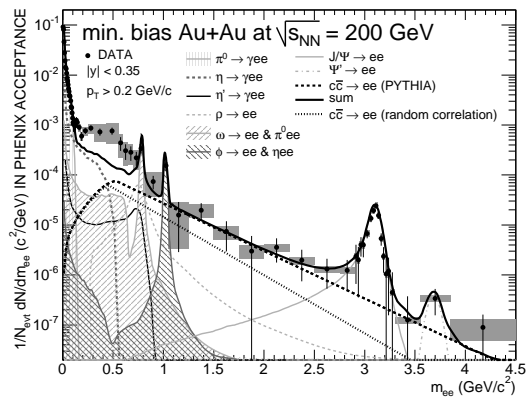}}
\vspace*{8pt}
\caption{The PHENIX low-mass spectrum for Au-Au collisions at top RHIC ~\cite{phenix_dileptons}.}  
\label{phenix_ee}
\end{figure}

The PHENIX experiment at RHIC has also
presented first results on di-lepton production in Au-Au
collisions~\cite{phenix_dileptons} at very high energy ($\sqrt{s_{NN}} = 200$
GeV). The results are presented in ~Fig.~\ref{phenix_ee}. In
addition to peaks of the vector mesons, one observes a very large enhancement
compared to the hadronic cocktail in the di-lepton yield for masses between 200
and 800 MeV. At present, the size of this enhancement is not reproduced
within the theoretical approaches described above. Future research
will tell whether new physics is visible here or whether these data can
also be described within the language of hadronic many-body theories.

Finally we would like to comment on the connection between the medium
modification of hadrons and the phase diagram. Already 25 years ago
Pisarski~\cite{pisarski} argued that prompt di-lepton production from a hot
fireball can be a signal for critical behavior through changes of the mass
and width of the $\rho$ meson near the phase boundary. The issue was further
investigated in~\cite{turko}. As discussed above, it is our current
understanding that near (or at) the deconfinement phase boundary also chiral
symmetry will be restored. The in-medium electromagnetic response, which is
dominated by the vector mesons $\rho, \omega$ and $\phi$, provides a direct
link to chiral symmetry and its restoration near $T_c$. Restoration of chiral
symmetry implies a strong reduction ('melting') of the quark condensate near
$T_c$. Furthermore, at the phase boundary, the vector- and axial-vector
correlation functions (see above) corresponding to the $\rho$ meson and its
chiral (parity) partner, the a$_1$ meson, must become identical in the limit of
vanishing quark masses~\cite{Weinberg,KaShu}.

It is an interesting observation~\cite{RaWa2} that the yield calculated using
the hadronic in-medium correlation function near the phase boundary coincides
remarkably well with that obtained from lowest-order $q \bar q$ annihilation
in the QGP, where chiral symmetry is restored. Since a strong increase of the
$\rho$-meson width is seen in the present di-lepton data, it thus seems that
the signal for chiral symmetry restoration in the electromagnetic response of
hot and dense matter is a smooth 'melting'~\footnote{Whether the melting occurs
only close to the phase boundary or over an extended range of temperatures
and densities below the phase boundary is currently an open issue.There are
indications, however, that already at chemical freeze-out the baryon density
is rather low~\cite{pbm_js_rho}.} of the $\rho$ meson into a featureless
quark-antiquark continuum (see also~\cite{kaempfer1_1, kaempfer1_2}). 
While this is not a rigorous argument for chiral symmetry restoration
by itself, a much stronger case could be made if the modification of the 'chiral' partner, 
the $a_1$ could be measured. On general grounds its spectral distribution  
has to become degenerate with that of the $\rho$ meson when 
chiral symmetry is restored. Hence, also the $a_1$ meson has to melt smoothly into
a quark-antiquark continuum. Unfortunately this is hard to check
experimentally since the dominant electromagnetic decay of the $a_1$ meson involves, 
besides a virtual or real photon, a pion. The latter suffers strong rescattering and absorption
in the fireball and hence the early stages of the collision are hard to probe.
At low temperatures and densities, however, one can prove
rigorously that chiral symmetry restoration manifests itself in a mixing of
the $\rho$- and $a_1$ meson through the absorption or emission of a pion from
the surrounding medium~\cite{ElIo}.

\subsection{Quarkonia--messengers of deconfinement}

In a nucleus-nucleus collision at very high energy heavy quarks, charm or
beauty, can be produced rather copiously. For example, the number of charm and
anti-charm quark pairs in a Pb-Pb collision at LHC energy might well reach
beyond a hundred. Because of the large mass of the charm quarks compared with
the typical QCD scale ($\Lambda_{QCD} \approx 200$ MeV, $m_{charm} \approx
1.3$ GeV) there is a separation of time scales between charm quark production
and the production of hadrons containing charm quarks~\cite{jpsi_medium}. The
question of medium modifications of such hadrons is then more subtle.

Particles collectively known as 'quarkonia' are bound states of charm or
beauty quarks and their antiquarks. They play an important role as probes for
deconfined matter inside the hot and dense fireball. In their seminal 1986
paper Matsui and Satz~\cite{satz} argued that the bound state made up of
charmed quarks and anti-quarks, the J/$\psi$ meson, would be destroyed (or
prevented from being formed) by the high density of partons in the QGP. The
physics behind this process is similar to Debye screening of the
electromagnetic field in an electromagnetic plasma through the presence of
movable electric charges. To provide a first estimate, we note that the density of
partons (quarks and gluons) in a non-interacting plasma with three massless flavors is $n
=4.2 T^3$. At a temperature of 500 MeV, this implies that $n\approx
70$/fm$^3$. The mean distance between these color charges scales like
$1/n^{1/3} \propto 1/T$ and is about 0.25 fm in the ideal gas limit, much
less than the spatial extent of the J/$\psi$ meson. Indeed, taking strong
interactions among the color charges into account leads to a ``Debye
screening'' radius $r_D \propto 1/(g_s(T) T)$ which decreases with increasing
temperature.  Hence the resulting color screening may destroy the bound state.
The suppressed yields of charmonia measured in a high-energy
nucleus-nucleus collision (compared to their production in the absence of a
QGP) was thus proposed~\cite{satz} as a 'smoking gun' signature for the QGP.

Measurements performed during the last decade at the CERN SPS accelerator
indeed provided first evidence for such a suppression~\cite{na50} in central
collisions between heavy nuclei. Little suppression was found in grazing
collisions or collisions between very light nuclei, where QGP formation is not
expected.  The precision data of \cite{na50} could be described, however, 
also by considering ``normal'' absorption of charmonium in the nuclear medium,
in conjunction with
its possible break-up by hadrons produced in the collision (co-movers).
Such mechanisms  could lead to charmonium suppression even in the
absence of QGP formation~\cite{gavin_vogt,capella,stoecker_jpsi} and the interpretation of the SPS
data remains inconclusive.

This situation took an interesting turn in 2000, when it was realized that the
large number of charm-quark pairs produced in a nuclear collisions at RHIC or
LHC energies leads to new mechanisms for charmonium production, either through
statistical production at the phase boundary~\cite{stat_rec_1, stat_rec_2}, or through
coalescence of charm quarks in the plasma~\cite{c_coal}. At low energy, the
mean number of charm-quark pairs produced in a collisions is much less than 1,
implying that a charmonium state, if at all, is always formed from charm quarks of
the one and only pair produced. On the other hand, the number of charm quark
pairs at RHIC energies is already much larger than 1 (indirect measurements
imply a charm-quark multiplicity of about 10) and the total number of charm
quarks in a collision at the LHC is expected to reach values larger than 100!
Under such conditions charm quarks from different pairs can combine to form
charmonium. Charm-quark recombination works effectively only if the charm
quarks can travel a significant distance in the plasma to 'meet' with their
prospective partner.  Under these conditions, charmonium production scales
quadratically with the number of charm quark pairs. Thus enhancement, rather
than strong suppression, is predicted~\cite{andr_06} for LHC energies. We note
that, in the recombination model, it is assumed that charmonia are either not
formed before the QGP or that they are completely destroyed by it (complete
quenching), so that all charmonium production takes place when the charm
quarks hadronize at the phase boundary. For a detailed discussion of this
point see~\cite{jpsi_medium}.

The most recent data from the RHIC accelerator provide interesting new insight into 
the connection between QGP formation and charmonium production but the
question how to use charmonia as messengers of deconfinement is far from
settled. Here we briefly describe the surprizing aspects of the new PHENIX
data and argue that they  lend first support to the
regeneration scenario described above. The major new insight came from a study
of the
rapidity\footnote{The 'rapidity' of a particle is defined through its total  
energy $E$ and the longitudinal momentum $p_z$ along the beam axis as:
$y=\frac{1}{2}\ln\left(\frac{E+p_z}{E-p_z}\right)$. In contrast to a particle's
velocity its rapidity $y$ is additive under Lorentz transformations.} and
centrality 
dependence (measured through the number of participating nucleons in the
collision) of the nuclear modification factor R$_{AA}^{J/\psi}$ which has, for the
first time, been measured by the PHENIX collaboration~\cite{phe1} in Au-Au
collisions.  This modification factor for $J/\psi$
production is defined as
\beq R_{AA}^{J/\psi}= \frac{\ud N_{J/\psi}^{AuAu}/\ud y}{N_{coll}\cdot\ud
N_{J/\psi}^{pp}/\ud y} 
\eeq
and relates the charmonium yield in nucleus-nucleus collisions to that
expected for a superposition of independent nucleon-nucleon collisions.  Here,
$\ud N_{J/\psi}/\ud y$ is the rapidity density of the $J/\psi$ yield for AA
and pp collisions and $N_{coll}$ is the number of binary nucleon-nucleon collisions for a
given centrality class.

\begin{figure}[ht]
\centering\includegraphics[width=.50\textwidth]{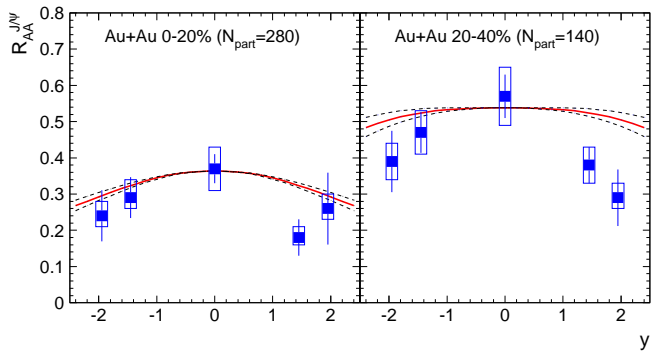}
\caption{Rapidity dependence of the nuclear modification factor
$R_{AA}^{J/\psi}$ for two centrality classes. The data~\cite{phe1} (symbols
with errors) are compared to calculations (lines). The dashed lines result
from uncertainties in the $J/\psi$ distribution in proton-proton
collisions.} 
\label{aa_fig1}
\end{figure}  

In Fig.~\ref{aa_fig1} we present the new data from the PHENIX
collaboration~\cite{phe1}. The most striking feature of these data is the
observation of a maximum in the rapidity dependence of $R_{AA}^{J/\psi}$ at
mid-rapidity (corresponding to y=0, i.e. production perpendicular to the beam
direction). This maximum was entirely unexpected as the observed trend is
opposite to that expected from the melting model~\cite{satz,satz1}, where
$R_{AA}^{J/\psi}$ should attain its smallest value (maximum suppression) in
regions of phase space with maximum energy density, i.e. near mid-rapidity.
Likewise, the destruction of charmonia by co-moving hadrons would also lead to
the largest suppression at mid-rapidity, in conflict with  PHENIX data.

\begin{figure}[hb]
\vspace{-.3cm}
\centering\includegraphics[width=.45\textwidth]{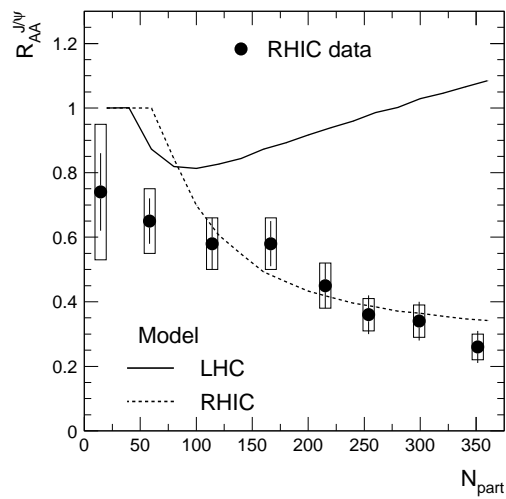}
\caption{Centrality dependence of the nuclear modification factor 
$R_{AA}^{J/\psi}$ for RHIC and LHC energies. The RHIC data are from the PHENIX
experiment~\cite{phe1}. The calculations are performed within the framework of
the statistical hadronization model~\cite{andr_06,andr_07}.} 
\label{aa_fig2}
\end{figure}

On the other hand, the observed maximum of $R_{AA}^{J/\psi}$ at midrapidity
is naturally explained in the recombination model of~\cite{andr_07} as being due
to enhanced charmonium production at the phase boundary:  the number of
charm quarks is maximal at mid-rapidity and this maximum is enhanced even further
through recombination. This mechanism provides  a good
description of the data, as indicated by the calculated curves in
Fig.~\ref{aa_fig1}. In this sense, the
PHENIX measurement constitutes first evidence for the
statistical production of J/$\psi$ at chemical freeze-out. Further support for
this interpretation comes from the observed centrality dependence of
$R_{AA}^{J/\psi}$ at midrapidity as shown in Fig.~\ref{aa_fig2}. We
reiterate that, if the
recombination model is correct, it implies complete charmonium
quenching in the QGP, as discussed above.

One should note that, at present, there are also other interpretations of the
PHENIX data, in particular through cold nuclear matter effects possibly 
reducing the number of gluons and,
hence, charm quarks when going away from mid-rapidity. Precision data on
$J/\psi$ production and its possible hydrodynamic flow will be needed to
distinguish between these different descriptions. The situation is concisely
reviewed in~\cite{GranierdeCassagnac:2008ke}. 

Since the number of charm quark pairs is still rather moderate at RHIC
energies, a strong enhancement of J/$\psi$ production is not expected in the
PHENIX measurement. The model predictions reproduce very well the decreasing
trend versus centrality seen in the RHIC data~\cite{phe1}. In contrast, at the
much higher LHC energy, the charm production cross section is expected to be
about an order of magnitude larger~\cite{rv1}. As a result, a totally opposite
trend as a function of centrality is predicted (Fig.~\ref{aa_fig2}) with
$R_{AA}$ exceeding 1 for central collisions. 

For these predictions it is assumed that charm quarks are effectively thermalized in the
very hot and dense QGP, implying that their recombination at the phase boundary gives
rise to a significant increase in yield near mid-rapidity. As shown, the
resulting predictions for measurements at LHC energies lead 
to a rather dramatic enhancement rather than suppression in
central Pb--Pb collisions. If observed, this would be a spectacular
fingerprint of a high-energy quark-gluon plasma, in which charm quarks are
effectively deconfined.  The data on charmonium production in Pb--Pb
collisions from the LHC will be decisive in settling the issue.

\section{Phases at High Baryon Density}

Nuclear matter can be compressed in two distinctly different ways: a rapid
squeeze that leads to strong heating, as realized in heavy-ion collisions at
relativistic energies; a slow squeeze which results in cold matter at very
high baryon density. This type of compression is impossible to achieve in the
laboratory but is  realized in the interior of a neutron star, a few
seconds after it has been born in a supernova explosion. It is conceivable
that, in the inner core, densities as high as ten times that in the middle of
a heavy nucleus can be reached~\cite{LaPr}. Under such conditions it is
expected that the closely packed neutrons (with a small admixture of protons,
electrons, and muons as well as baryons carrying strange quarks) dissolve into
their constituents and the u,d,s-quarks form a degenerate Fermi
liquid\footnote{Because of their much larger (Higgs) masses charm, bottom and
top quarks play no role at the relevant densities.}. The composition is
determined by charge and color neutrality and the requirement of $\beta$
equilibrium, i.e. equilibrium of weak interaction processes.

It has long been known that fermionic systems at low temperatures become
unstable to the formation and condensation of 'Cooper pairs' if the
interaction between two fermions is attractive. This situation is expected in
quark-gluon matter above the deconfinement
transition~\cite{Barrois,Frautschi_1}. Here the Cooper instability of the
Fermi surface and the formation of di-quark pairs is mediated by the
attractive interaction induced by gluon exchange between two quarks of
specific color, flavor and spin combinations. Since such combinations carry
net color, the new state is called a 'color superconductor' a term that was
first used in refs.~\cite{Barrois,Frautschi_1}. The presence of color
superconductivity in the core of neutron stars could  
lead to interesting new effects in the long-time evolution of such objects
such as modifications of the cooling rate through neutrino emission,
instabilities caused by gravitational wave radiation of pulsars or glitches in
the spin-down rate~\cite{ASRS}.

\subsection{Color Superconductivity}

\begin{figure}[th]
\centerline{\includegraphics[width=0.45\textwidth]{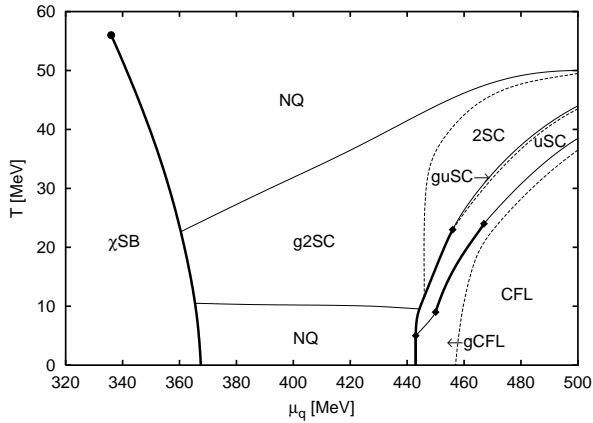}}
\vspace*{8pt}
\caption{Color superconducting phases at high baryo-chemical potential $\mu_q
= \mu_b/3$  and low
temperatures as predicted by the NJL model in the Hartree
approximation~\cite{Ruest}. The region of spontaneously broken chiral
symmetry is denoted by $\chi$SB while regions where quark matter is in the
normal state are indicated by NQ. The bold solid lines mark boundaries of
first-order transitions, while the thin lines denote second-order
boundaries. The dashed lines indicate the boundaries between gapless and
gapped regions. Several critical points are found.}  
\label{csc_phases}
\end{figure}

Early analysis of the possible pairing patterns in cold quark matter and
estimates of the resulting gaps $\Delta$ based on the exchange of a single
gluon~\cite{Bailin_1,Bailin_2} led to values of a few MeV for $\Delta$.
Such low values have little influence on the high density EoS. This situation
changed in the 1990's when it was pointed out that, in the physically
interesting region of $\mu_q\sim 500$ MeV which corresponds to about ten times
the density in a heavy nucleus, perturbative one-gluon exchange 
is inadequate because of the strong increase in $\alpha_s$ at such momentum
scales. The resulting non-perturbative effects in the quark-gluon coupling
were estimated in the NJL model and led to gap values of up to 100
MeV~\cite{Alford_SC,Rapp_SC}. Subsequently, it was found that the many
possible combinations of flavor-color and spin degrees of freedom, dictated by
the fermionic antisymmetry of the Cooper pair wave function, can lead to a rich
phase structure~\cite{ASRS,Rischke,Buballa}. For total spin $S=0$ one has the
possibility of pairing two quark flavors, say up or down, leaving the third
flavor unpaired (the socalled '2SC' phase) or all three quark flavors can
participate. In this case there is a definite combination of color and flavor
degrees of freedom called 'color-flavor locking' (CFL). Under the conditions
of charge and color neutrality as well as $\beta$ equilibrium the Fermi
energies of quarks with given color and flavor quantum numbers are in general
not equal. The imbalance is partly caused by the mass difference
$m_s-m_{u,d}$. For a large mismatch, pairing with unequal quantum numbers
becomes difficult ('stressed superconductivity') and can even lead to
'gapless' phases (g2SC~\cite{ShoHu}, gCFL~\cite{AlKouRaja}). Also 
crystalline phases similar to the LOFF phases~\cite{FF,LO} in conventional
superconductors are 
conceivable~\cite{ASRS}. Which phase is favored at a given temperature and
density is determined by the global minimum of the free energy. An example is shown
in Fig.~\ref{csc_phases}. Comparing Fig.~\ref{phases_ex} and
Fig.~\ref{csc_phases} it is unlikely that any of the high-density and
low-temperature phases can be explored in heavy-ion collisions.

Stressed superconductivity can however be studied experimentally in trapped
ultracold fermionic atomic gases~\cite{Giorgini}. Here an imbalance in
chemical potentials can be achieved by populating two hyperfine states of the
atom with a different number of particles. At the same time the interaction
strength can be controlled using Feshbach resonances, to drive the system from
weak coupling 
(BCS regime) through the point where diatomic bound states form to the point 
where diatomic molecules undergo Bose-Einstein condensation (BEC regime). Thus
many of the predicted phases of cold quark matter can be 'simulated' in the
laboratory with interesting future perspectives and cross-fertilization.

Even though the NJL model is useful in exploring the many possibilities of
superconducting phases, its quantitative predictive power is limited by the
large sensitivity of the results to the model parameters. First principle
calculations, on the other hand, are very difficult since they require
accurate knowledge of the di-quark interaction on scales of the Fermi energy
$\epsilon_F$ where $\alpha_s$ is large. Only at very high densities or
asymptotically large $\mu_q$, the coupling becomes small enough to make
reliable predictions from first principles. In this case one-gluon exchange
between di-quarks dominates. In the dense medium its longitudinal
(color-electric) component is Debye screened while the transversal
(color-magnetic) components are dynamically screened due to Landau damping.
This implies that the ratio of the magnetic to electric polarization functions
goes like $\omega/|\vec q|$ where $\omega$ is the frequency and $\vec q$ the
three-momentum of the gluon field. In the static limit $\omega\to 0$ the
magnetic components therefore remain unscreened. As a consequence, in contrast
to the usual BCS theory where the pairing gap as a function of the coupling
constant $g$ varies as $\Delta/\mu\sim\exp(-const/g^2)$, one has~\cite{Son} 
\beq
\frac{\Delta}{\mu_q}\sim\exp{\left(-\frac{3\pi^2}{\sqrt{2}g_s}\right)}~.
\label{CSC}  
\eeq
Such retardation effects for long-range forces are also known in condensed
matter physics~\cite{Eliashberg_1,Eliashberg_2}. The $1/g_s$-dependence in the exponent of the
gap function leads to the surprizing phenomenon that the pairing gap can take
arbitrarily large values, even though the coupling decreases~\cite{RajaWil}\footnote
{Note that $g_s=\sqrt{4\pi\alpha_s}$ according to Eq.~(\ref{alphas_run}) behaves
like $\sqrt{1/\ln\mu_q}$ if one assumes that the momentum scale $Q$ is governed by
$\mu_q$. Inserting this into Eq.~(\ref{CSC}) it is clear that the exponential drops more 
slowly than $1/\mu_q$.}. Taking into account the color-flavor-spin degrees of freedom one 
finds the CFL phase to be the energetically most favored pairing state at
asymptotically large quark chemical potentials (Fig.~\ref{csc_phases}).

Even though these ab-initio findings are quite interesting from a many-body
point of view, they are valid only for asymptotically large values of $\mu_q$,
because of the logarithmic running of 
$\alpha_s$ (Eq.~\ref{alphas_run})\footnote{For weak coupling theory to apply in QCD
$\mu_q$ has to be of the order of $10^{4}$ MeV! (Fig.~\ref{alpha_s}).}. Hence, they are of little
relevance for the interior of neutron stars where $\mu_q\sim 400-600$ MeV. One
can try to remedy this by the inclusion of higher-order corrections in $g_s$.
Since at such scales $g_s\simeq 1$ it is questionable, however, whether such
perturbative expansion schemes are justified. A more promising approach is to
use Schwinger-Dyson equations where both the quark and gluon fields are
treated non-perturbatively with a proper treatment of infra-red
(small-momentum behavior) of $\alpha_s$. Recent
results~\cite{Nickel_1,Nickel_2} indicate that, in the relevant regime of
quark densities in the core of neutron stars, pairing gaps of the order of 100
MeV can be expected, confirming the earlier findings within NJL model studies.

\section{Summary and Conclusions}

In the 30 years since the first discussions about the phases of QCD and the
corresponding phase diagram there has been tremendous progress in our
understanding of strongly interacting matter at extreme conditions. Large
experimental campaigns have been mounted and have amassed a wealth of new data
and led to a series of discoveries. Here we have concentrated on aspects
relevant to the QCD phase diagram. In particular, we have discussed that for
symmetric matter $(\mu_b=0)$ the chemical freeze-out temperature can be
determined with an uncertainty of better than $10 \%$ from measured hadron
abundances. We have further argued that the observed temperature behavior lends
strong support to the notion of a critical temperature $T_H$ introduced by
Hagedorn and provided arguments that $T_H$ coincides with $T_c$, the critical
temperature for the quark-hadron transition of strongly interacting matter.
Thus an important point in the phase diagram has been established
experimentally. We have furthermore summarized the evidence for mass changes
of hadrons near the phase boundary, with particular emphasis on the $\rho$
meson and laid out arguments how these findings are connected to the
restoration of chiral symmetry near the phase transition line. Charmonium
production is apparently also strongly influenced by the QCD phase transition
and we have outlined the particular role of this production process for
studies of deconfinement.  Along with the experimental progress also came
impressive theoretical developments, both concerning ``exact'' solutions of
QCD on a discrete space-time lattice as well as the development of powerful,
effective models to study the physical processes emerging from the
experimental observations.

What may be expected in the future?  With the experimental
program at RHIC and in particular the heavy-ion program at the CERN LHC
\footnote{The LHC is now scheduled to start operations in the summer of 2009, first with
protons and afterwards with a pilot run for the Pb beam program.} the structure
of the matter above $T_c$ and at vanishing chemical potential can be studied
quantitatively. In particular, the fireballs formed in Pb-Pb
collisions at LHC energies will have much higher initial temperatures, maybe
reaching 1 GeV, and live much longer ($>$10 fm lifetime up to the
quark-hadron phase transition) than those produced at RHIC. Furthermore,
hard probes, in particular high transverse-momentum jets and heavy quarks, will be
abundantly produced. From studies in this new environment should emerge not
only detailed tests of ab-initio QCD predictions about the phase transition as
well as information about the bulk properties of the QGP at high temperature
and its stopping power for high momentum quarks
but also insight into the nature of the processes that lead to confinement.
Studies of the phases of strongly interacting matter at high densities and
moderate temperatures, on the other hand, are still in their infancy. The development of relevant
effective theories (including the complex reaction dynamics) as well as
developments of lattice QCD simulations at finite chemical potentials are
important milestones in the understanding of quark matter at high densities.
Further experimental studies at lower energy at the RHIC collider as well as
with the planned CBM experiment at the FAIR facility at GSI are mandatory to
make progress in our understanding of the QCD phase transition in the high
density regime.

\begin{acknowledgments}
We thank A. Andronic, M. Buballa, K. Redlich and A. Richter for a critical reading of the manuscript. 
\end{acknowledgments}


\end{document}